\documentclass[sigconf]{acmart}
\AtBeginDocument{%
  \providecommand\BibTeX{{%
    \normalfont B\kern-0.5em{\scshape i\kern-0.25em b}\kern-0.8em\TeX}}}

\copyrightyear{2023}
\acmYear{2023}
\setcopyright{rightsretained}
\acmConference[CI '23]{Collective Intelligence Conference}{November 6--9, 2023}{Delft, Netherlands}
\acmBooktitle{Collective Intelligence Conference (CI '23), November 6--9, 2023, Delft, Netherlands}\acmDOI{10.1145/3582269.3615595}
\acmISBN{979-8-4007-0113-9/23/11}

\makeatletter
\gdef\@copyrightpermission{
  \begin{minipage}{0.2\columnwidth}
   \href{https://creativecommons.org/licenses/by-nc-sa/4.0/}{\includegraphics[width=0.90\textwidth]{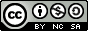}}
  \end{minipage}\hfill
  \begin{minipage}{0.8\columnwidth}
   \href{https://creativecommons.org/licenses/by-nc-sa/4.0/}{This work is licensed under a Creative Commons Attribution-NonCommercial-ShareAlike International 4.0 License.}
  \end{minipage}
  \vspace{5pt}
}
\makeatother

\usepackage{graphicx} 
\usepackage{subcaption}
\usepackage{makecell}
\usepackage{tabularx}

\usepackage{colortbl}

\usepackage[suppress]{color-edits}
\usepackage{color-edits}
\addauthor{ak}{blue} 
\addauthor{lg}{cyan} 
\addauthor{hz}{green} 
\addauthor{kh}{magenta} 
\addauthor{sc}{red}
\addauthor{ml}{red}
\addauthor{placeholder}{red}

\addauthor{khold}{black} 
\addauthor{lgold}{black} 
\addauthor{akold}{black} 
\addauthor{hzold}{black} 

\begin{document}

\title[Training Towards \textit{Critical Use}]{Training Towards \textit{Critical Use}: Learning to \\Situate AI Predictions Relative to Human Knowledge
}

\author{Anna Kawakami}
\affiliation{%
  \institution{Carnegie Mellon University
  \city{Pittsburgh}
  \state{PA}
  \country{USA}}
}
\email{akawakam@cs.cmu.edu}

\author{Luke Guerdan}
\affiliation{%
  \institution{Carnegie Mellon University}
  \city{Pittsburgh}
  \country{USA}}
\email{lguerdan@cs.cmu.edu}

\author{Yanghuidi Cheng}
\affiliation{%
  \institution{Carnegie Mellon University}
  \city{Pittsburgh}
  \country{USA}}
\email{ycheng8610@gmail.com}

\author{Matthew Lee}
\affiliation{%
 \institution{Toyota Research Institute}
 \city{Los Altos}
 \state{CA}
 \country{USA}}
\email{matt.lee@tri.global}

\author{Scott Carter}
\affiliation{%
 \institution{Toyota Research Institute}
 \city{Los Altos}
 \state{CA}
 \country{USA}}
\email{scott.carter@tri.global}

\author{Nikos Arechiga}
\affiliation{%
 \institution{Toyota Research Institute}
 \city{Los Altos}
 \state{CA}
 \country{USA}}
\email{nikos.arechiga@tri.global}

\author{Kate Glazko}
\affiliation{%
  \institution{University of Washington}
  \city{Seattle}
  \country{USA}}
\email{glazko@cs.washington.edu}

\author{Haiyi Zhu}
\authornote{Both co-senior authors contributed equally to this research.}
\affiliation{%
  \institution{Carnegie Mellon University}
  \city{Pittsburgh}
  \country{USA}}
\email{haiyiz@cs.cmu.edu}

\author{Kenneth Holstein}\authornotemark[1]
\affiliation{%
  \institution{Carnegie Mellon University}
  \city{Pittsburgh}
  \country{USA}}
\email{kjholste@cs.cmu.edu}

\renewcommand{\shortauthors}{Anna Kawakami et. al.}


\begin{abstract}
A growing body of research has explored how to support humans in making better use of AI-based decision support, including via training and onboarding. Existing research has focused on decision-making tasks where it is possible to evaluate ``appropriate reliance'' by comparing each decision against a ground truth label that cleanly maps to both the AI's predictive target and the human decision-maker's goals. However, this assumption does not hold in many real-world settings where AI tools are deployed today (e.g., social work, criminal justice, and healthcare). In this paper, we introduce a process-oriented notion of appropriate reliance called \textit{critical use} that centers the human's ability to situate AI predictions against knowledge that is uniquely available to them but unavailable to the AI model. To explore how training can support critical use, we conduct a randomized online experiment in a complex social decision-making setting: child maltreatment screening. We find that, by providing participants with accelerated, low-stakes opportunities to practice AI-assisted decision-making in this setting, novices \khedit{came}\khdelete{learned} to exhibit \khdelete{similar }patterns of disagreement with AI \khedit{that resemble those of}\khdelete{as} experienced workers. A qualitative examination of participants' explanations for their AI-assisted decisions revealed that they drew upon \khedit{qualitative case narratives}\khdelete{case specific allegation information}, to which the AI model did not have access, to learn when (not) to rely on AI predictions. Our findings open new questions for the study and design of training for real-world AI-assisted decision-making. 

\end{abstract}



\maketitle

\section{Introduction} \label{Introduction}
\authornote{Both co-senior authors contributed equally to this research.}
AI-based decision support (ADS) tools are increasingly \akdelete{used} \akedit{deployed to assist frontline professionals} in critical, real-world contexts like social services, healthcare, and criminal justice, with the hope of improving \akdelete{human }decision quality. However, realizing these tools’ potential in practice is far from guaranteed~\cite{buccinca2020proxy,dietvorst2015algorithm,Green2019,holstein2022designing,lee2004trust}. To support \akedit{human decision-makers} \akdelete{frontline professionals }in effectively \akdelete{and responsibly }using such tools, it is critical that they receive sufficient training before using them in practice. Yet, today, ADS tools are often introduced into these contexts without adequately \akdelete{onboarding and training for}\akedit{preparing} the frontline professionals who are asked to use them day-to-day~\cite{cai2019hello,kawakami2022improving,Saxena2021}. 

\akedit{To address this real-world need, a growing body of studies have explored \khedit{onboarding and training} approaches to support human decision-makers in learning to use ADS tools, \khedit{demonstrating promising initial results}\khdelete{including through new onboarding and training approaches}~\cite[e.g.,][]{bansal2019beyond,mozannar2021teaching}.  \khdelete{These studies have so far demonstrated that training interventions can help yield more accurate joint human-AI decisions.}}
\akedit{ 
Many of these research studies state a motivation to improve human-AI decision quality in critical applications (e.g., healthcare, criminal justice, and social services). However, a careful review of these studies surfaces key differences \khedit{between their task designs versus}\khdelete{between the properties of AI-assisted decision-making assumed in the study versus} the actual AI-assisted decision-making tasks frontline professionals perform in their day-to-day work~\cite{guerdan2023ground}:
\khedit{
\begin{itemize}
    \item \textbf{Target-construct mismatch:} Whereas most experimental studies of AI-assisted decision-making assume that human experts and AI models are predicting the same construct, in practice 
    models are often trained to predict an \khedit{imperfect proxy} for the true outcome of interest to human decision-makers~\cite{wang2022against,guerdan2023ground,coston2022validity,jacobs2021measurement,kleinberg2018human}. In these settings, effective use of AI predictions depends critically on humans' ability to account for such misalignment.
    \item \textbf{Information asymmetry:} While experimental studies typically assume that humans and AI models have access to the same information for decision-making, in practice humans frequently have access to \textit{complementary} information~\cite{hemmer2022effect,holstein2020conceptual,kawakami2022improving,holstein2023toward}. In these settings, effective use of AI predictions depends critically on humans' ability to make use of decision-relevant information to which they have unique access.
\end{itemize}
Although target-construct mismatch and information asymmetry are often discussed in studies of real-world AI-assisted decision-making, their implications for human-AI interaction design remain underexplored. It is an open question how people can best be supported in learning to make AI-assisted decisions in complex, real-world contexts.
}
}

\akedit{In this paper, we explore these questions} in a complex, real-world domain \khedit{involving both properties described above:} \khedit{AI-assisted} child maltreatment screening\khedit{. The}\khdelete{, a setting in which the} use of AI-based decision support tools \khedit{in child welfare} is rapidly spreading~\cite{aclu2021family,saxena2020human}. \khedit{Given significant target-construct mismatch and information asymmetry, it is not advisable in this context}\khdelete{In these sorts of critical, real-world applications of ADS, it may not be advisable} to train humans to make decisions that \khedit{perfectly} align with the imperfect proxy used by the AI model~\cite{cheng2022child,De-Arteaga2020,kawakami2022improving}. 
\khedit{For the same reason, it is unclear how to apply existing measures of ``appropriate reliance'' on AI predictions in this context, given that these typically assume that alignment with model's target is the goal.} 
\khedit{To address this gap, we define the notion of}\khdelete{\akedit{Instead, we explore what ``appropriate reliance'' means in these real-world contexts through defining a notion of}} \textit{\textbf{critical use}}---\scedit{that is, humans' ability to situate \akedit{AI predictions against potentially complementary knowledge uniquely available to them (but not the AI model). } }

\khedit{Adopting \textit{critical use} as a lens, we}\khdelete{Using supporting\textit{critical use} as our training goal, we scope our study to understand the effects of providing decision-making practice and showing decision feedback. In particular, we } conducted a within-subjects, randomized controlled experiment to \khedit{investigate \textit{what} and \textit{how} people learn through practice and feedback on decision-making, in the context of AI-assisted child maltreatment screening.} 
\akedit{
\khedit{We designed}\khdelete{To address these research questions, we design} training activities that \khedit{simulate the actual tasks workers face in a real-world deployment context---Allegheny County's use of the Allegheny Family Screening Tool (AFST)~\cite{Chouldechova2018,eubanks2018automating,kawakami2022improving}---but provide participants with accelerated opportunities to practice making AI-assisted decisions, low-stakes setting.}\khdelete{provide opportunities to practice making AI-assisted child maltreatment screening decisions on decision tasks that have historically been presented to actual frontline professionals in a real-world deployment context.} 

\khedit{To inform our study design and analysis, we drew upon an extensive body of findings from prior field studies in the AFST context, examining how experienced workers make AI-assisted decisions in this setting~\cite{cheng2022child,cheng2022heterogeneity,chouldechova2018case,De-Arteaga2020,kawakami2022improving}. Prior literature indicates that experienced workers in this context use the AFST's predictions in sophisticated ways, which improve decision-making. For example, quantitative analyses of workers' decision-making over a multi-year period found that, when workers exercised their discretion to disagree with model predictions on a case-by-case basis, this has the aggregate effect of reducing errors and racial disparities in decision-making, compared with what would have happened had they uncritically agreed with the model's predictions~\cite{cheng2022child,De-Arteaga2020}. In line with our definition of \textit{critical use}, field observations revealed that workers drew upon their knowledge of information unavailable to the AFST (e.g., relevant context from qualitative case narratives) to calibrate their reliance upon the model's predictions in each case~\cite{kawakami2022improving}. To support analysis of study participants' learning, we leveraged our knowledge of experienced workers' AI-assisted decision-making practices to define \khdelete{a set of }\textit{indicators} of critical use in this domain. 
}
} 

\akedit{Overall, our analyses suggest that our training activities \khedit{promoted more critical use of AI predictions}\khdelete{supported \textit{critical use}}. \akedit{In earlier practice opportunities, participants in both conditions started off relying on AI model predictions more heavily\khdelete{, including}\khedit{. They did so even} on cases for which the model \khedit{erred with respect to}\khdelete{failed to accurately predict} the proxy it was trained on. However, w}\akdelete{W}ith increased practice, participants learned to disagree with the AI predictions more often\akdelete{\khedit{, and}\khdelete{. At the same time, participants} began relying more upon diverse information sources \khedit{to inform their decisions} (e.g., details from qualitative case narratives)\khdelete{, to justify their AI-assisted decisions}}. 
\khedit{In particular,}\khdelete{ \akedit{through these patterns of disagreement,}} participants\khdelete{With increased practice, participants simultaneously } learned to make\khdelete{Receiving repeated practice opportunities led participants to make} decisions that were more accurate \khedit{with respect to the}\khdelete{w.r.t.} decision-making of actual experienced workers.}
\akedit{The training also improved participants'}\khedit{\akdelete{Interestingly, although providing explicit feedback on participants' decisions improved their} ability to predict the AI model's behavior. \akedit{Interestingly, participants who received practice alone saw greater improvements in their ability to predict model behavior, compared to participants who also received explicit feedback alongside practice. Moreover, providing explicit feedback} did not impact their learning with respect to decision-making, compared with the effects of repeated practice alone. Our analyses indicate that participants used the qualitative case narratives as a powerful form of \textit{implicit} feedback on the reliability of individual AI predictions, through which participants learned to calibrate their reliance on the AI system. This suggest that, beyond exploring mechanisms for \textit{explicit} feedback on human- and AI-based decisions, future work on training for AI-assisted decision-making should explore the design of rich, implicit feedback mechanisms that empower learners to cross-check AI predictions against concrete, qualitative representations~\cite[cf.][]{stampfer2011eliciting}.
}
More broadly, our findings open up a space of new opportunities and open questions for the design of effective training materials and evaluation metrics for AI-assisted decision-making.

This paper makes the following contributions: 
\begin{itemize}
    \item  \khedit{We introduce the \textbf{concept of \textit{critical use}}, which can be applied to AI-assisted decision-making settings where traditional measures of ``appropriate reliance,'' based on available ground truth labels, are likely to be unreliable.}
    As a learning goal, critical use is applicable to a range of real-world contexts that involve target-construct mismatch and information asymmetry (e.g., healthcare, criminal justice, content moderation, and education). We explore \textbf{how critical use can be measured} via a case study in the child maltreatment screening context. 



    \item  We present the \textbf{first experimental investigation} in the literature of \textit{what} and \textit{how} humans learn through practice and feedback in settings where human knowledge complements that of an AI model (through target-construct mismatch and information asymmetry). Through our analyses, we investigate \textbf{how critical use can be supported through training}.
    %

    \item Based on our findings, we highlight \textbf{new opportunities and open questions} for the design of effective training materials and learning measures in real-world AI-assisted decision-making settings.
    
\end{itemize}

\section{Background and Research Questions} \label{Background}
\khedit{In this section, we first briefly overview prior literature on training for AI-assisted decision-making, and then discuss key opportunities to foster human-AI complementarity that are under-explored within this literature. Finally, we motivate and present our research questions.}\khdelete{\akedit{First, we overview  \hl{fill in first section content.} Next, we discuss prior work on \hl{fill in second section content.} For each subsection, we describe how our study is grounded in and extends upon existing literature.}} 
\begin{figure*}
    \includegraphics[width=\textwidth]{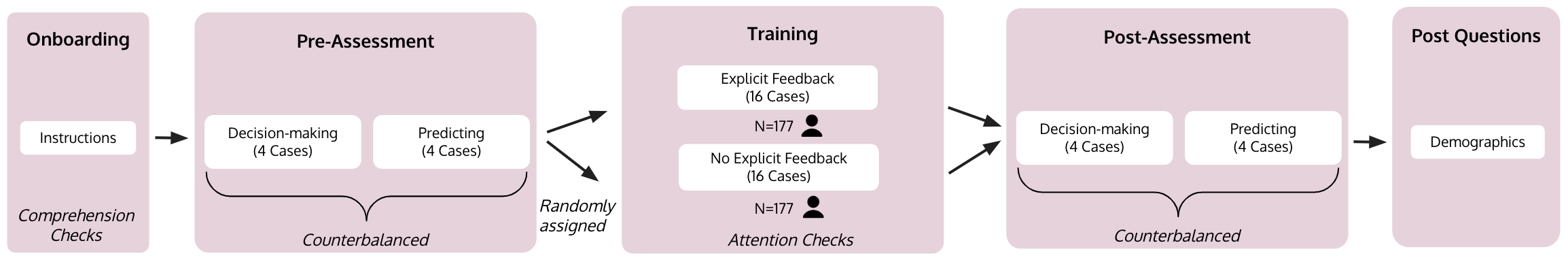}
    \caption{A high-level overview of the study design. The study included five phases: Onboarding, Pre-Assessment, Training, Post-Assessment, and Post Questions. In the Training phase, participants were randomly assigned between-subjects to the \textit{Practice} or the \textit{Practice + Explicit Feedback} case.}
     \label{fig:study_overview}
\end{figure*} 
\subsection{\khedit{Training for AI-Assisted Decision-Making}}
Recent years have seen the rapid adoption of AI-based decision support (ADS) tools to assist human decision-making in settings like social services, education, healthcare, and criminal justice~\cite{barocas-hardt-narayanan,Chouldechova2018,Levy2021,yu2018artificial}. 
\khdelete{While AI-based decision support (ADS) tools are designed to help overcome limitations in human decision-making, a growing body of literature has documented ways in which ADS tools have failed to adequately support the frontline professionals who are asked to use them day-to-day\kholdedit{, or have even introduced new burdens and challenges into their work}~\cite{HoltenMoller2020,holstein2017intelligent,kang2022stories,kawakami2022improving,Saxena2021,sendak2020real,veale2018fairness,yang2016investigating}.} 
Despite \kholdedit{rapid growth in adoption}, frontline professionals are often asked to use these tools day-to-day without adequate training. Even in high-stakes contexts, where AI-assisted decisions have the potential to change the trajectory of individuals’ lives, the implementation of \khdelete{improved }training approaches has severely lagged behind investments in \khedit{model development}\khdelete{the ADS model itself}~\cite{cai2019hello,cai2021onboarding,kawakami2022improving,Saxena2021}.  \akedit{This lag has also been recognized in recent regulatory and policy efforts, for example, through government directives requiring employee training on AI-based decision-making tools~\cite{secretariat2020directive}.}

To \kholdedit{address this gap, an emerging line of research in HCI and CSCW has begun to explore the design of onboarding and training interventions to} support humans in learning to \kholdedit{work effectively with ADS tools}. For example, Cai et. al. conducted a qualitative study with pathologists to identify their needs and desires for \kholdedit{onboarding to human-AI collaborative decision-making}. \kholdedit{Other research has begun to investigate the effectiveness of different training approaches via online experiments.} For example, Monzannar et. al. \kholdedit{introduced an exemplar-based interactive training approach}\kholddelete{explored strategies for selecting teaching cases} to hone human decision-makers’ mental models of a question answering model’s strengths and weaknesses, \kholdedit{and explored strategies for selecting training examples that could help humans learn an AI model's error boundaries most efficiently}~\cite{mozannar2021teaching}. Lai et. al. explored model-based tutorials as an intervention to improve humans' abilities to understand patterns of model behavior, using a case study in deceptive hotel review detections~\cite{lai2020chicago}. Overall, \kholdedit{this line of work has shown promising early results, suggesting that} the proposed training approaches can support human decision-makers in learning to make more accurate AI-assisted decisions. 

However, prior studies on training for AI-assisted decision-making have \kholdedit{focused on relatively well-defined decision tasks}. In the next section, we discuss properties of real-world AI-assisted decision-making \khedit{that have potential to support human-AI complementarity, but which are not typically represented in experimental study designs.}\khdelete{often missing in existing human subjects AI-assisted decision-making studies (including in training), and their potential for supporting learning towards human-AI complementarity.}

\subsection{Potential Sources of Human-AI Complementarity in Decision-Making}

While there is an expectation that both human and AI-based judgements are imperfect, there is hope that careful integrations of the two can improve decision-making by drawing upon the ~\textit{complementary} strengths of each~\cite{bansal2021does,zhang2022,donahue2022human,holstein2018student, kawakami2022improving,wilder2020learning}.  Prior literature on measurement in AI-assisted decision-making~\cite{kleinberg2018human,guerdan2023ground,rastogi2022unifying} has articulated \khedit{several properties of real-world decision-making tasks that can support to human-AI complementarity in practice. In this paper we focus closely on two of these:}\khdelete{two (amongst other) properties, which we posit is consequential to the design of appropriate training materials}: 
\akedit{ 
\begin{itemize}
    \item \textbf{Target-construct mismatch}\khdelete{~\cite{wang2022against} (a.k.a. target misalignment~\cite{coston2022validity}, outcome measurement error~\cite{guerdan2023ground})}: \khedit{In many real-world deployment settings, AI models are trained to predict \textit{imperfect proxies} for the true construct of interest to decision-makers~\cite{coston2022validity,guerdan2023ground,jacobs2021measurement,kleinberg2018human,wang2022against}. }
    For example, in \khedit{healthcare}\khdelete{AI-assisted clinical decision-making}, \khedit{a widely-deployed ADS tool was trained to predict}\khdelete{ADS tools may predict} \textit{medical costs} as a proxy for clinicians' actual decision-making goal of assessing \khedit{patients'} \textit{medical need} ~\cite{mullainathan2021inequity}. \lgedit{\khedit{However, an analysis found that medical}\khdelete{Medical} cost \khedit{is a systematically worse proxy for medical need among certain demographic groups---for instance,}\khdelete{is an imperfect proxy because} hospitals may have historically turned patients away from care due to a lack of insurance or other discriminatory factors \cite{obermeyer2019dissecting}}. 
    \item \textbf{Information asymmetry}\khdelete{~\cite{holstein2020conceptual,hemmer2022effect} (a.k.a. presence of ``unobservables''~\cite{kleinberg2018human})}: \khedit{In real-world settings, humans and AI models frequently have access to complementary sources of information~\cite{holstein2020conceptual,kawakami2022improving,kleinberg2018human}.}
    \khdelete{Prior work has also discussed how frontline professionals in real-world decision settings have access to different, often qualitative and case-specific information sources which is unobservable to the AI model.}
    For example, 
    \akdelete{\kholdedit{in the context of}\kholddelete{literature on} AI-assisted \kholddelete{decision-making tools in} healthcare decision-making,}
    clinicians \khdelete{may }make decisions about medical resource allocation across high-risk patients\akdelete{with chronic illness} 
    by drawing upon a patient's physical presentation\lgedit{, subjective assessments of their well-being, and real-time test results\khedit{---information that may be unobservable to an AI model~\citep{mullainathan2019machine}}.} \lgdelete{or subjective assessment of their well-being}.
\end{itemize}
} 

%
Beyond the healthcare examples provided above, these two properties are observed across a wide range of other real-world \khedit{AI-assisted decision-making settings, including}\khdelete{ADS applications:} child welfare, criminal justice, online content moderation, education, lending, and hiring, to name a few. 
\kholdedit{Recent field research has shown that, in some settings, frontline}\kholddelete{Frontline} professionals \kholddelete{using ADS in social decision-making settings}are aware of \khedit{information asymmetries and}\khdelete{ the} misalignments between their own \kholdedit{goals as human decision-makers versus the}\kholddelete{ and} \kholdedit{proxy outcomes that an AI model is trained to predict\khedit{.}\khdelete{, and furthermore that this} \khedit{Furthermore, this} awareness can shape how they calibrate their reliance on AI predictions on a case-by-case basis}\kholddelete{an AI model’s decision targets}~\cite[e.g.,][]{cheng2022child, holstein2022designing, kawakami2022improving}. 
\khdelete{Prior research has also hypothesized that the qualitative, contextual information sources unobservable to the AI model may be a means through which humans and AI can achieve complementary performance~\cite{}. \akdelete{Across a range of domains--from risk recidivism~\cite{} and child welfare~\cite{} to content moderation~\cite{} and education~\cite{}--these two properties \hl{....}.}} 
Yet \khedit{to date,} most existing \khdelete{human subjects }\khedit{experimental research on AI-assisted decision-making has constructed study environments where these properties are absent.} 
In fact, in \khedit{a recent review of the literature}\khdelete{recent work}, Guerdan et. al.~\cite{guerdan2023ground} found that 92\% of \khedit{experimental studies}\khdelete{human subjects studies for human-AI decision-making} \khedit{make the assumption that no}\khdelete{exclude} target-construct mismatch \khedit{or}\khdelete{and }information asymmetries \khedit{are present}.


Prior studies exploring ways to support humans in learning to make AI-assisted decisions involve tasks such as identifying defective objects \cite{bansal2019beyond}, passage-based question answering \cite{mozannar2021teaching}, comparing nutritional content \cite{gajos2022people}, or identifying deceptive hotel reviews \cite{lai2020chicago}. In each of these decision tasks, there is no information asymmetry: the human decision-maker is  only provided access to the same or a subset of the type of information the AI model is trained on. There \khedit{is also no pronounced}\khdelete{also exists no} target-construct mismatch: the AI model used in the task is trained on a predictive target (e.g., accuracy of AI-assisted decisions about the \textit{nutritional content} of a given meal \cite{gajos2022people}, or whether an \textit{object is defective} \cite{bansal2019beyond}) that directly corresponds to both the \khedit{AI and the human's}\khdelete{AI's} predictive target\khdelete{ and the human's underlying decision-making goal}. Like other studies evaluating AI-assisted decision-making~\cite{lai2021towards}, accuracy and learning (i.e., changes in accuracy over time) is assessed via correspondence to this ground truth signal. 
\akdelete{However, existing notions of how ``good'' a given AI-assisted decision is often focuses on outcome-based assessments. Calibrating appropriate reliance--\hl{insert definition}--..... \hl{...include 1-2 sentences about under- and over-reliance....}}

While prior work \khedit{indicates}\khdelete{has hypothesized} that \khedit{the presence of target-construct mismatch and information asymmetries presents potential for human-AI complementarity in real-world settings}\khdelete{these two information sources (target-construct mismatch and information asymmetries) may be a means through which human decision-makers can more appropriately rely on AI predictions, and thus achieve complementary performance}, we know little about how to \khedit{help humans learn to leverage this potential.}\khdelete{\akedit{leverage this potential to} support human\khedit{s}\khdelete{ decision-makers} in learning to use ADS tools in real-world settings.} In this paper, we begin to explore this opportunity space. 
\subsection{Research Questions}
Domains involving human decision-making have long recognized that \textbf{realistic practice} and \textbf{tailored feedback} are key processes for learning. Literature from the learning sciences has demonstrated how practice improves performance across domains, from academic subjects like math and psychology to complex medical procedures and even to professionals' abilities to navigate challenging social interactions~\cite{helsdingen2011effects,kaka2021digital,koedinger2012knowledge}. In prior literature, these simple practice effects have been substantially enhanced when supplemented with \akedit{explicit} feedback~\cite{el2010factors,koedinger2012knowledge,koedinger2007exploring}. 
\kholdedit{In real-world decision-making settings such as social work, healthcare, and education, where there is typically a long lag (on the order of months or even years) between decisions and corresponding outcomes, such simulation-based training approaches have the potential to greatly accelerate learning.}

In our study, we \khedit{investigate \textit{what} and \textit{how} people learn through repeated}\khdelete{simulate child maltreatment screening decisions by providing repeated, realistic practice opportunities} \khedit{practice making AI-assisted decisions in the context of child maltreatment screening.} \khedit{In addition, we explore the effects on participants' learning}\khdelete{and explore the additional effects on participants' learning} of showing explicit feedback on \khedit{their}\khdelete{participants'} decisions. \khedit{We}\khdelete{Using this training approach, we} ask the following research questions:
\begin{enumerate}
    \item Can \textbf{repeated practice} making AI-assisted screening decisions help promote \khedit{more} critical use of AI predictions? 
    \item Does providing \textbf{explicit feedback} on \khedit{decisions}\khdelete{, in the form of multiple, complementary feedback signals} help promote critical use of AI predictions, compared to \khdelete{repeated }practice alone?
\end{enumerate}

\section{Methods} \label{Methods}
To \khedit{investigate our research questions}\khdelete{explore how people learn to make AI-assisted decisions \kholddelete{through practice and feedback, }in settings involving 
target-construct mismatch and information asymmetries}, we designed and conducted a randomized \kholddelete{behavioral }online experiment with crowd workers, social work graduate students, and social workers. 

\subsection{Context} \label{Context}

We \kholdedit{ground our investigations in the}\kholddelete{focus on the} context of \kholddelete{AI-assisted }child maltreatment screening: a complex, social decision-making context where the use of AI-based decision support tools is rapidly spreading~\cite{aclu2021family,saxena2020human,Chouldechova2018}. The task design and data are \kholddelete{specifically }drawn from \kholdedit{a real-world ADS} deployment\kholdedit{: the Allegheny Family Screening Tool (AFST).} \kholddelete{context of} \kholdedit{As one of the}\kholddelete{the Allegheny Family Screening Tool ( AFST} \kholdedit{longest deployed and most well-known ADS tools in social services, }\kholddelete{), an ADS tool that}\kholdedit{the AFST} has influenced the design \kholdedit{and deployment} of many other AI-assisted decision-making tools across the U.S.~\cite{aclu2021family}. The Allegheny County Department of Human Services (DHS) deployed the AFST in 2016 to assist hotline call screeners and supervisors in \kholddelete{assessing the risk and }prioritizing \kholddelete{among referred }alleged child maltreatment cases \kholdedit{for investigation}~\cite{Chouldechova2018}. As is common for ADS tools deployed in complex, real-world settings, the AFST \khedit{predicts readily measurable yet indirect \textit{proxies}}\khdelete{involves the use of imperfect \textit{proxies}} for the \khedit{construct of interest to decision-makers}\khdelete{actual goals of human decision-makers}. \kholddelete{In the context of the AFST and child maltreatment screening more broadly, the gap between AI models’ predictive targets and human goals can be quite large:}\mldelete{While}\mledit{Whereas} frontline decision-makers focus on \kholdedit{ensuring children's} immediate safety\kholddelete{and short-term risks}, the AFST predicts \khedit{a proxy outcome}\khdelete{\kholdedit{a readily-measurable, yet indirect \textit{proxy} for safety}}\kholddelete{outcomes} on a much longer time horizon: a child's risk of \khedit{being \textit{placed into foster care }}\khdelete{\textit{out-of-home placement within two years}}\khedit{within the next \textit{two years}} ~\cite{cheng2022child,kawakami2022improving}. \kholdedit{The AI-assisted child maltreatment context}\kholddelete{This AI-assisted decision-making context} also involves information asymmetries: \khedit{prior field studies find that }frontline \khedit{workers' decisions are informed by rich, qualitative information about a given case (e.g., reported allegations), which is unavailable to the ADS model}\khdelete{use rich, case-specific details from the reported allegations---information that is not accessible to the ADS model---to inform their decisions and interpret the model’s predictions}~\cite{kawakami2022improving,saxena2022unpacking}.

In their day-to-day work, call screeners and supervisors at Allegheny County are asked to use the AFST, which outputs a risk score on a scale from 1 (low risk of \khdelete{out-of-home }placement) to 20 (high risk of \khdelete{out-of-home }placement). \khedit{Our study activities are designed to simulate}\khdelete{The design of \kholdedit{our study activities is directly}\kholddelete{the training tasks are} inspired by} the actual decision process that call screeners complete. The AI-assisted decision-making task at Allegheny County begins once the call screener receives a call (e.g., from a teacher or neighbor) reporting \khedit{alleged child maltreatment}\khdelete{on an alleged child maltreatment case}. \khedit{The call screener takes notes on the call, and then}\khdelete{During this time, the call screener writes notes on the allegations they are hearing from the caller. After hearing this allegation, the call screener} uses an online data system to collect relevant administrative records on the family being reported. Finally, \kholdedit{these administrative records are used}\kholddelete{the call screener inputs the collected administrative records }to generate the AFST score. Using information from the allegations, \khdelete{collected }administrative records, and the AFST score, the call screener then makes a recommendation about whether \khedit{or not} to screen in \khdelete{or screen out }the \khedit{family}\khdelete{child} for investigation. Then, the \khdelete{call screening }supervisor makes the final screening decision, informed by the recommendation from the call screener, administrative records, and the AFST score. \kholddelete{Prior literature has discussed various limitations of the AFST model and challenges social workers encounter when using the AFST in practice, along with opportunities to improve the use of the model by empowering social workers [with more information about the model]~\cite{AFSTdocumentation,kawakami2022improving}. }We describe how \khedit{our study simulates this task in the next subsections.}\khdelete{the training interfaces simulate this real-world decision-making }\khdelete{subsection~\ref{Study Design}, and detail the training data drawn from historical records from this context in subsection~\ref{case_selection}.}  

\subsection{Study Design and Materials} \label{Study Design}
%
\kholdedit{To investigate our research questions, we conducted an online experiment with a total of 354 participants, using a pretest-postest experimental design. An overview of our study design is shown in Figure~\ref{fig:study_overview}. Following a brief \khedit{onboarding}\khdelete{\textbf{onboarding}} to the study, participants first completed a brief \textbf{pre-assessment}, which provided a baseline assessment of their knowledge and decision-making \khdelete{behavior} prior to our experimental intervention. Participants then proceeded to a \textbf{training phase}, during which they were randomly assigned to practice making AI-assisted decisions either \textit{with} or \textit{without} explicit feedback. \khedit{Finally, participants}\khdelete{Participants then} completed a \textbf{post-assessment}, which was structurally identical to the pre-assessment, but presented participants with new, unique cases. \khedit{At the end of the post-assessment, participants were asked a small}\khdelete{Finally, participants completed a} set of \khedit{post-study questions}\khdelete{\textbf{post-study questions}} \khedit{aimed at understanding}\khdelete{intended to better understand} their backgrounds.} \akedit{The IQR for study completion time is [48 minutes, 1 hour 25 minutes] with a median of 1 hour 4 minutes.}


In the \textbf{training phase}, participants \kholdedit{practiced making AI-assisted decisions on a series of}\kholddelete{completed} 16 practice cases\kholdedit{, presented in random order}\kholddelete{ for the AI-Assisted Decision-Making activity}. Participants were randomly assigned between-subjects to one of two conditions: \textit{Practice} or \textit{Practice + Explicit Feedback}. \kholdedit{As described in the subsections below, for}\kholddelete{For} each of the 16 cases shown in the Training phase, \kholdedit{participants in the \textit{Practice} condition were asked to review the allegations, administrative records, and the AI model's risk score, and then make a screening decision. Participants in this condition were shown the next case immediately after inputting their decision. By contrast, after making a screening decision,} participants in the \textit{Practice + Explicit Feedback} condition \kholddelete{were asked to make a screening decision, then }were \kholdedit{presented with} \kholddelete{given }immediate feedback on their decision. \kholdedit{After seeing the feedback, participants were then asked to indicate whether they would have made the same decision, \textit{``knowing what you know now''}, or whether they would change their decision based on the feedback they saw.} \kholddelete{Participants in the \textit{Practice} condition were simply asked to make a screening decision, then were shown the next case after inputting their screening decision.} \khdelete{\akedit{Because participants in each condition first completed the pre-assessment before entering the training phase, we did not include a control condition.}} 

\kholdedit{The \textbf{pre- and post-assessments}} were designed to assess both (a) how participants make child maltreatment screening decisions with AI assistance; and (b) participants' ability to predict what risk score the AI model will assign to a given case. \kholdedit{Assessing participants' ability to predict AI outputs, both}\kholddelete{By assessing participants' ability to predict the AI risk score} before and after the training phase\kholddelete{, we aimed to assess} \kholdedit{enabled us to investigate} whether repeated practice and feedback on AI-assisted decision-making would additionally improve their ability to \kholdedit{mentally simulate the AI model's behavior~\cite{buccinca2020proxy,lipton2018mythos,Poursabzi-Sangdeh2021}}\kholddelete{predict the decisions of an AI model that is trained on an imperfect proxy}. \kholdedit{Each assessment included a series of}\kholddelete{Therefore, we include} four AI Score Prediction activities and four AI-Assisted Decision-Making activities \kholdedit{(which were identical to the decision-making activities presented during the training phase)}. \kholddelete{within each Pre-Assessment and Post-Assessment}\kholdedit{During the assessment phases of the study, participants were not shown any explicit feedback on either their \khdelete{AI-assisted }decisions or their AI score predictions, regardless of which experimental condition they were assigned to for the training phase.} In each assessment, we counterbalanced the order in which the two types of assessment activities were shown. For each case shown during the assessment phases, participants were asked to elaborate (via open text) on how they made their AI-assisted decision or AI score prediction immediately after inputting their response.

\kholdedit{Below, we describe the designs of the two types of activities included across the training and assessment phases.}
\begin{figure}
    \includegraphics[scale=0.25]{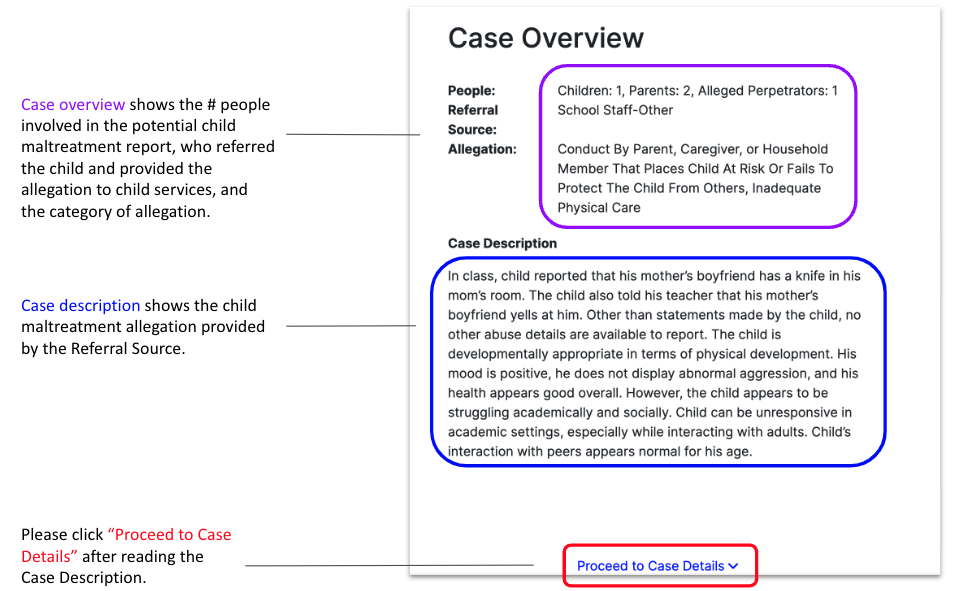}
    \caption{A screenshot of the Case Overview information panel that provides an overview of the alleged child maltreatment case and a text box \mledit{with the allegations as a qualitative description}. The case content, including the allegations, are taken from real historical data recording past experienced workers' AI-assisted decision-making tasks.}
     \label{fig:case_overview}
\end{figure} 

\begin{figure}
    \includegraphics[scale=0.25]{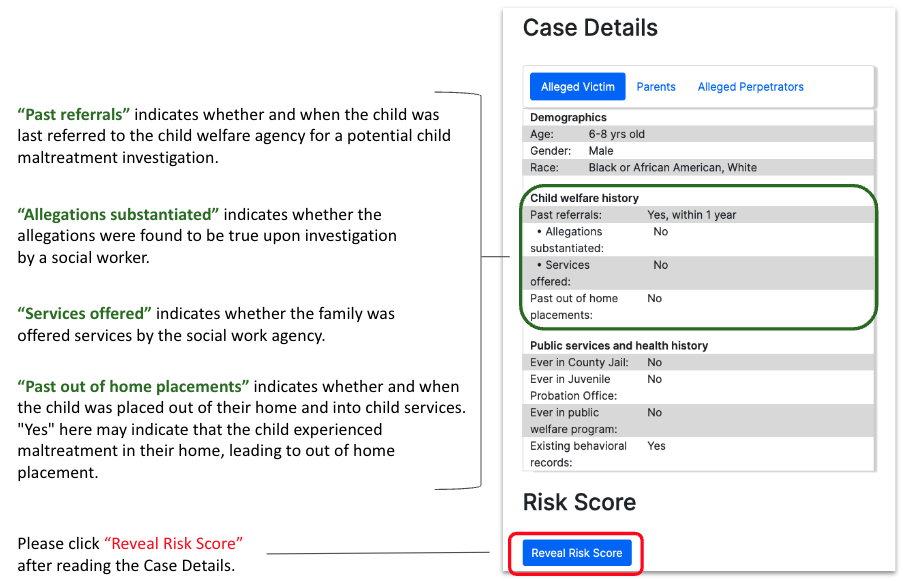}
    \caption{A screenshot of the Case Details information panel that provides public records and past referral information on the individuals in the alleged child maltreatment case. The content is taken from the same historical case information used to populate the Case Overview and AI Risk Score.}
     \label{fig:case_details}
\end{figure}
\begin{figure}
    \includegraphics[scale=0.25]{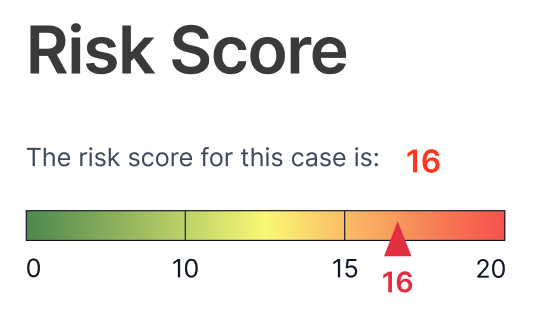}
    \caption{A screenshot of the AI Risk Score \mledit{interface} shown to participants in the AI-assisted decision-making task. \mledit{This} mirrors the actual AFST interface \mledit{used by} social workers.}
     \label{fig:risk_score}
\end{figure}

 
\paragraph{AI-Assisted Decision-Making Activity}
To inform their decisions in the AI-assisted decision-making activity, participants were presented with both a referral vignette and an AI prediction. For each case, the AI prediction and content in the referral vignette was \kholdedit{drawn and adapted} from \khedit{case notes on} actual cases \khedit{for which past workers subsequently}\khdelete{on which past social workers} made AI-assisted decisions\kholdedit{, in the Allegheny County context}. The referral vignette first shows a \textbf{Case Overview} which includes \textbf{qualitative allegations}. The child maltreatment allegations were typed by experienced workers based on calls they received (e.g., from a neighbor or a teacher). For each allegation drawn from the historical dataset, we removed any potentially identifying information (e.g., details pertaining to specific regions or households), while still preserving high-level properties of the allegations, as \kholdedit{this information could be relevant to the interpretation of}\kholddelete{they could influence how a human interprets} other case-related information, including the AI prediction\kholddelete{, or how likely it is to observe certain downstream outcomes of a case}. Figure~\ref{fig:case_overview} shows a screenshot of a \textbf{Case Overview }section for a sample referral vignette. Next, we showed \textbf{Case Details} (Figure~\ref{fig:case_details}) including demographic information, child welfare history, and public services and health history of the alleged child victim, parent(s), and alleged perpetrator(s). 
\khdelete{In practice, actual workers may have access to a wider range of demographic information for a given case.}For the purposes of our experiment, we opted to show a smaller subset of demographic information \khedit{than actual workers are able to access in practice}, both to exclude potentially identifying information and to ensure that participants would have sufficient time to look through the information (cf.~\cite{Cheng2021}). 
Finally, we showed the \textbf{Risk Score }(Figure~\ref{fig:risk_score}): the actual AFST score for the \khedit{given} historical case\khdelete{ represented by the Case Overview and Case Details}. 

\begin{figure}
    \includegraphics[scale=0.3]{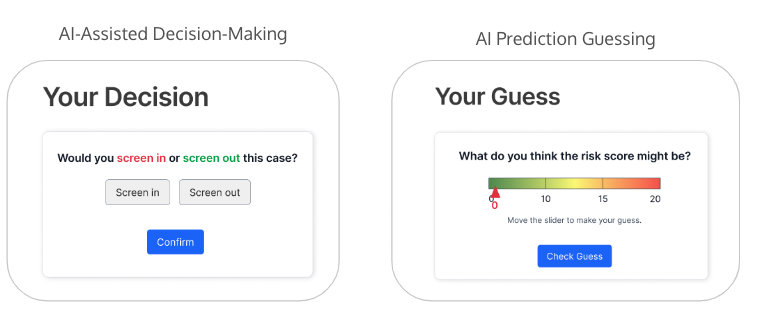}
    \caption{A screenshot of the question that participants are asked in the AI-assisted decision-making activity (left) and AI score prediction activity (right). To inform this decision, participants also see information from the Case Overview, Case Details, and (for the decision-making task only) the AI prediction.}
     \label{fig:dm_guess}
\end{figure}

\kholdedit{The presentation order of the Case Overview, Case Details, and Risk Score}
\kholddelete{Taken together, participants were shown the Case Overview, Case Details, and AI Prediction, in that order, before being asked to make a screening decision (shown in Figure~\ref{fig:dm_guess} [left]). The ordering} was designed to reflect the actual order in which social workers encounter each information source \kholdedit{in the AFST deployment context, when reviewing a case.} 
We designed the training interface to gradually reveal this information\kholdedit{, requiring participants to click ahead in order to reveal the next information source}.\kholddelete{The user needed to click on buttons (e.g., ``Proceed to Case Details’’) to see the next information source, to encourage the user to read each information source in the order in which it is presented.} 

\paragraph{AI Score Prediction Activity}
In the AI Score Prediction activity, participants were similarly shown referral vignettes (including the Case Overview and Case Details) that replicated the decision-making tasks of actual social workers using an ADS. However, rather than showing the AI Risk Score, participants were asked to make their own prediction on the AI output on a scale of 1 to 20. 
\kholddelete{We include the AI Score Prediction activity in the Pre- and Post-Assessments (described further in [section name]) to assess correlations between how accurately the participant can guess the AI output and how they learn to make AI-assisted decisions through practice and feedback.} 
Figure~\ref{fig:dm_guess} (right) includes a screenshot of the question participants were asked after seeing the referral vignette.

\subsection{Feedback Design}

\begin{figure}
    \includegraphics[scale=0.235]{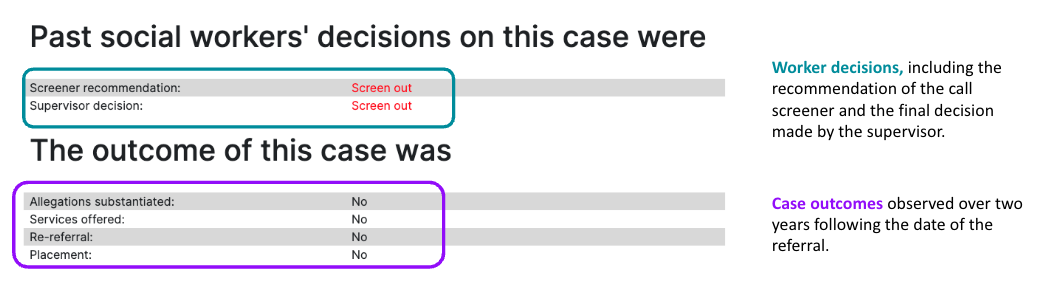}
    \caption{A screenshot of the six imperfect proxies (screener recommendation, supervisor decision, allegation substantiation, services offered, re-referral, and placement) shown as feedback to participants in the AI-assisted decision-making task \mledit{in the \textit{Practice + Explicit Feedback} condition}. Participants in the \textit{Practice} condition were shown the next case immediately after submitting their decision. 
    }
     \label{fig:feedback}
\end{figure}

Participants in the \textit{Practice + Explicit Feedback} condition received \kholdedit{immediate} feedback on their decisions during the study's training phase. 
\khdelete{In complex, social decision-making settings such as child maltreatment screening, it is typically the case that neither humans nor AI models have access to a single, agreed upon source of ``ground truth,'' either at training or decision time. Instead, both humans and AI models must rely on incomplete information for learning and decision-making (e.g., \textit{indirect evidence} from allegations or \textit{imperfect proxies} for the true outcomes of interest).} Given that \khedit{we do not typically have access to a}\khdelete{there is no} single\khedit{, reliable}\khdelete{\kholdedit{observed}\kholddelete{observable and incontestable}} ``ground truth’’ \khedit{label} in the context of child maltreatment screening, \khedit{as discussed above,} we \khdelete{instead }explore the learning effects of providing participants with multiple feedback signals in two primary forms: \textit{\kholdedit{observed}\kholddelete{observable} outcomes} and \textit{historical human decisions} (Figure~\ref{fig:feedback}).
\khdelete{Training humans to make decisions that only align with \textit{\kholdedit{observed}\kholddelete{observable} outcomes} that are recorded in available datasets (such as whether an allegation was substantiated, or whether there was an out-of-home placement within two years) carries several potential risks. For example, humans might learn to align their decisions with simplistic proxies rather than more meaningful decision-making criteria (cf.~\cite{green2021algorithmic}). In turn, humans might learn to replicate harmful biases in outcome data (e.g., due to selection and intervention effects~\cite{coston2022validity,kleinberg2018human}). On the other hand, training a human to make decisions that only align with \textit{past frontline professionals' decisions} would rest on the faulty assumption that all past workers made perfect decisions. There would be no opportunities to learn from mistakes of the past. In this experiment, we consider that feedback on \kholdedit{observed}\kholddelete{observable} outcomes (which are uniquely available only retrospectively, for training-like purposes) could serve as a complementary source of information to help human learners interpret the accuracy or desirability of the past experienced workers’ decisions.}
By showing multiple, imperfect ``ground truths,’’ we provide the human decision-maker with opportunities to reflect on disagreements amongst the ground truths and their own initial decisions, to learn how to more appropriately calibrate reliance on the ADS model over time.

Feedback in the form of \textit{\kholdedit{observed}\kholddelete{observable} outcomes} \kholdedit{that were available}\kholddelete{, selected based on availability} in the historical administrative dataset, included \kholdedit{the following.} For each, we briefly summarize its limitations as a feedback signal: 
\begin{enumerate}
    \item Whether the child was \textbf{placed within two years}: \kholdedit{In the AFST context and in our study, t}\kholddelete{T}his is the AI model’s predictive target. If the child was eventually placed out of their home within two years, this may signal that the child was maltreated and should have been screened in and subsequently investigated. However,\scdelete{as documented in prior literature,} there are other unobservable factors unrelated to the screening decision that may have led to this outcome~\cite{coston2022validity}. For example, the child’s parents may suddenly pass away\scdelete{in a car crash}, requiring the child to be placed into foster care in the absence of any reliable grandparents or relatives. Thus, placement in two years is \khedit{an imperfect}\khdelete{not a perfect} proxy for the accuracy of the screening decision.  
    
    \item Whether the child was \textbf{re-referred to the agency within two years}: Similar to (a),\mldelete{a child being re-referred to the child maltreatment agency within two years may indicate that the child was maltreated and s hould have been screened in. \textit{However}, once again, \scdelete{prior literature documents}} a range of other unobservable factors unrelated to the decision \scdelete{that} may have led to this outcome. For example,\scdelete{prior literature has documented how} certain stakeholders (e.g., divorced parents doing ``retaliation calls’’) may feel incentivized to call the hotline and alleged maltreatment, when in fact, the child is safe ~\cite{kawakami2022improving}.

    \item Whether the \textbf{allegations were substantiated upon screen in}: Only cases that were historically screened in can observe this outcome. Allegations of potential child maltreatment being substantiated based upon investigation may indicate that the child was indeed maltreated, and thus screening in was the ``right’’ decision. \textit{But},\scdelete{the allegation itself, even if substantiated, may not necessarily indicate that the child was maltreated and should have been investigated. For example,} allegations, even if substantiated, may result from cultural misunderstandings or implicit biases from the caller \lgdelete{. Prior literature has suggested these caveats may lead to increased calls and screen ins for children in historically marginalized communities~}\cite{cheng2022child}. 
    \item Whether \textbf{services were offered to the family upon screen in}: Only cases that were historically screened in can observe this outcome. A family that is offered services may indicate that the child was maltreated, and thus was offered services that can support recovery. In this case, a decision to screen in would be ``right.’’ \textit{On the other hand,} even if a child was not maltreated, a family may receive services as a preventive measure to minimize risk of future maltreatment.  
\end{enumerate}

Feedback in the form of \textit{historical human decisions} included: 
\begin{enumerate}
    \item Whether the \textbf{call screener’s recommendation }was a screen in or screen out: This is the call screening recommendation a past experienced social worker made, based on the case information and AI risk score they saw. 
    \item Whether the \textbf{call screening supervisor’s final decision }was to screen in or screen out: This is the final decision that a past experienced social worker made, based upon a review of case information, the AI risk score, and the call screener's recommendation. 
\end{enumerate}

\khdelete{Our feedback signals, like the information shown in the Case Overview, Case Details, and AI Prediction, were drawn from the actual historical decisions made by past experienced social workers who were assigned a given case, along with the actual eventual outcomes of those social workers’ decisions.} \kholdedit{Taken together, this set of feedback signals captures both the accuracy of the AI model’s predictions (with respect to the proxy outcome it is trained to predict), as well as what past experienced human workers believed was the right decision after reviewing both the case information and the AI prediction.}
\kholddelete{To train humans on human-AI complementarity, we provide feedback that signal both the accuracy of the AI model’s predictions, as well as what past experienced human workers believed was the ``right’’ decision after reviewing the case information. In particular, participants were shown two sets of imperfect proxies:\textit{ \kholdedit{observed}\kholddelete{observable} outcomes }and \textit{historical human decisions} (Figure~\ref{fig:feedback}).} 


\subsection{Case Selection}\label{case_selection}
The \kholdedit{cases shown in the} training and assessment activities are informed by the historical administrative records of referrals at Allegheny County between June and December of 2019. \lgdelete{This time period coincides with deployment of the AFST Version 2, which is the most recent version of the tool for which complete referral data was available.} \kholdedit{In this section, we describe how cases sampled and distributed across phases of the study.}
\kholddelete{The data used to populate the training interface were drawn from a combination of referral-level metadata (Figure <case/overview>, Figure <feedback>), written case notes containing allegation details recorded by call screeners during referral processing (Figure <case/details>), and feature-level data used to generate risk scores (Figure <case/details>). The AI Risk Score interface mirrors the actual AFST interface that social workers in Allegheny County see.}  

\subsubsection{Case Categorization} \label{CaseCategorization}

We developed four high-level case categories (and eight sub-categories) to \kholdedit{support the systematic sampling of historical cases for}\kholddelete{the cases to draw the historical administrative data from to} use in our \kholddelete{experimental }study. \lgdelete{we developed four high-level case categories (and eight sub-categories) that reflect variations along the supervisor decisions\kholdedit{,} \kholddelete{and }AI predictions\kholdedit{, and recorded case outcomes}.} 
\kholddelete{Our case categories are defined over the following dimensions: 
\begin{itemize}
    \item \textbf{AI Prediction: }We designate cases with an AFST score $\geq$ 15 as ``screen in’’ and cases with an AFST score $<$ 15 as ``screen out’’. This screen in threshold \kholdedit{aligns with the high risk cutoff shown to call screeners on the AFST tool interface \cite{AFSTdocumentation}. Use of this threshold also} follows prior analyses of automated AFST decisions (Cheng, Stapleton et al., 2022)\kholddelete{and also coincides with the high risk cutoff shown to call screeners on the AFST tool interface \cite{}}.
    \item \textbf{Past Supervisor Decision: }Historical worker decisions were determined based on the supervisor's final decision for a referral. 
    \item \textbf{Case Outcome: }Our primary welfare outcome used for case categorization involves whether a child was placed outside the home in foster care within two years of the referral. We adopt this outcome because it is the AFST prediction target and has been used extensively as a measure of decision quality in prior research \lgoldedit{\cite{chouldechova2018case}.}
\end{itemize}
}
\lgedit{Our taxonomy for selecting cases consists of the following alignment-measures:} 
\lgdelete{Based on concordance between these three dimensions, \kholddelete{(i.e., the AI Prediction [screen in/out], Past Supervisor Decision [screen in/out] and Case Outcome [placed/not placed])} we constructed three alignment measures:} 
\begin{itemize}
    \item \textbf{Worker-AFST alignment: }\khdelete{This}\khedit{Cases where}\khdelete{measure is\textit{ true} if} the worker agreed with the AFST decision\footnote{We designate cases with an AFST score $\geq$ 15 as ``screen in’’ and cases with an AFST score $<$ 15 as ``screen out’’. This screen-in threshold \kholdedit{aligns with the high risk cutoff shown to call screeners on the AFST tool interface \cite{AFSTdocumentation}. Use of this threshold also} follows prior analyses of automated AFST decisions \cite{cheng2022child}\kholddelete{and also coincides with the high risk cutoff shown to call screeners on the AFST tool interface \cite{}.}} (i.e., both state screen in or screen out).\khdelete{and \textit{false} otherwise.} 
    \item \textbf{Worker-outcome alignment: }\khedit{Cases where}\khdelete{This measure is\textit{ true} when} the worker's assessment \kholdedit{aligns}\kholddelete{coincides} with the observed outcome (i.e., screen in \& placed or screen out \& not placed). \khdelete{and \textit{false} otherwise. }
    \item \textbf{AFST-outcome alignment:} \khedit{Cases where}\khdelete{This measure is\textit{ true }when} the AFST's assessment \khdelete{of risk }\kholdedit{aligns}\kholddelete{coincides} with the observed outcome.\khdelete{(i.e., screen in \& placed or screen out \& not placed). }\khdelete{and \textit{false} otherwise.} 
\end{itemize}

\begin{figure}
    \centering
    \includegraphics[scale=0.5]{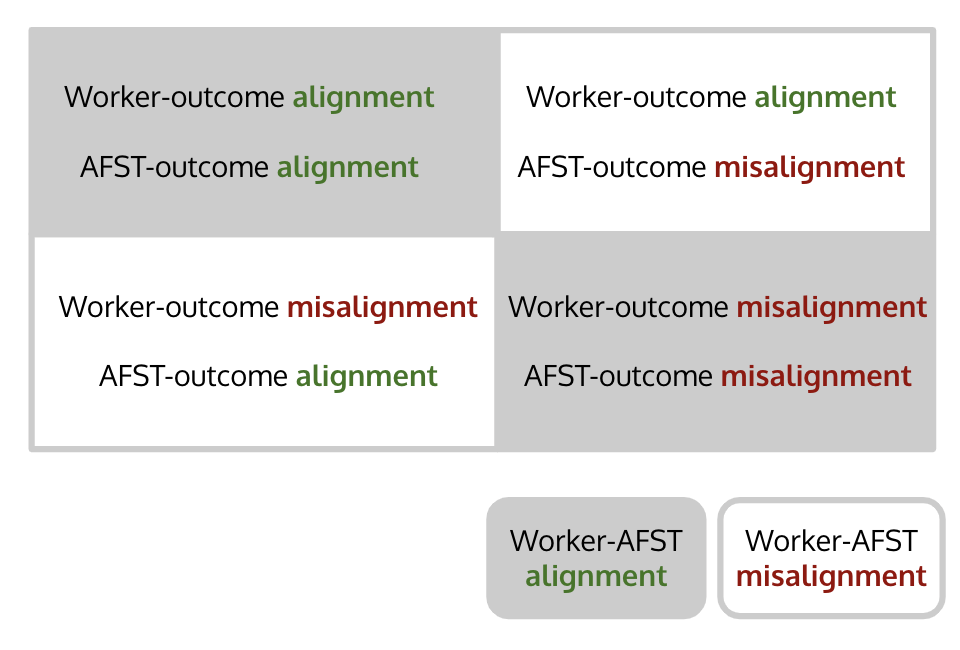}
    \caption{A 2x2 confusion matrix breaking down the three alignment measures into four high-level categories. }
     \label{fig:matrix}
\end{figure}

To guide the sampling of cases, we then constructed a 2x2 confusion matrix breaking down the three alignment measures into four categories as shown in Figure~\ref{fig:matrix}. \lgdelete{The top \kholdedit{left}\kholddelete{right} panel \kholdedit{of this figure} contains cases in which the worker and AFST decision are both aligned with the observed outcome. The bottom right panel contains cases in which the worker and AFST decision are both misaligned with the observed outcome. This main diagonal (shaded in gray) also reflects Worker-AFST alignment, in which both the human and model agreed in assessments of risk. Conversely, the off-diagonal (in white) represents cases in which the worker and AFST disagreed historically, where one or the other was aligned with the outcome.}

\subsubsection{Case Sampling and Randomization}

We \lgedit{randomly sampled cases from historical referral data according to our case categorization.} We drew 6 cases from each of the 8 sub-categories, yielding a total of 48 cases. \lgdelete{Historical referral data was segmented \kholdedit{according to our case categorization}\kholddelete{into the eight high-level case categories outlined above}, and six cases from each category were randomly sampled to include in our case pool. For each case sampled for our experiment, we manually checked the allegation details associated with the case to ensure the quality of the documentation. If the allegations documented were sparse or difficult to interpret, we instead randomly selected another case mapping to the same case category.}Cases were sampled with equal weighting across categories to enable \kholdedit{us to study}\kholddelete{studying} learning effects \kholdedit{relative to}\kholddelete{as a function of} historical \kholddelete{case }worker and algorithm \kholdedit{behavior}\kholddelete{performance}. \lgdelete{Therefore, Worker-AFST, Worker-outcome, and AFST-outcome alignment were each true for 50\% of cases \kholdedit{shown to participants}. This resulted in a total of forty-eight cases.}Cases within each category were then randomly assigned across the pre-assessment, training, and post-assessment phases of the study. Across participants, cases were assigned to the same study phase (i.e., pre-assessment, training, or post-assessment) to enable item-level analyses, but were presented in randomized order within each phase.

\subsection{Participants and Recruitment}
\kholddelete{Aside from the feedback condition, along with additional demographic questions for participants with social work background, all participants were shown identical information in the study.} 
\kholddelete{Participants were compensated \$13 for completing the study. Participants who also had social work domain expertise were compensated an additional \$7 for a total of \$20.} 
We recruited a total of 354 participants \kholdedit{through the crowdsourcing platform Prolific, social media\kholddelete{channels (Facebook \kholdedit{and}\kholddelete{groups,} LinkedIn groups)}, and direct email.}\kholddelete{, including 177 participants each in the \textit{Practice} and \textit{Practice + Explicit Feedback} conditions.}
\kholddelete{All participants were based in the U.S. \kholddelete{,}\kholdedit{and were} above 18 years of age\kholddelete{, and had never taken the study before}.}To explore \kholdedit{potential impacts of domain expertise}\kholddelete{the effects of having domain expertise in social work}, we \kholdedit{sought to include}\kholddelete{included} participants with \kholdedit{domain knowledge in} social work\kholddelete{expertise} in our participant pool. 
\kholdedit{First, we reached out via email to}\kholddelete{On email, we directly reached out to} social work professors at US-based universities and directors of US-based \kholddelete{non-profit}social work \kholdedit{non-profits}\kholddelete{organizations}, and asked them to share \kholdedit{our study invitation}\kholddelete{the online study} with social work graduate students and/or practicing social workers in their networks. 
\kholddelete{We \kholddelete{additionally}used social media channels and direct email to recruit participants with social work domain expertise.} 
\kholdedit{Second, we advertised the study on}\kholddelete{On Facebook and LinkedIn, we created postings to} relevant social work and child welfare groups such as ``Social Work Network'' and ``Child Welfare, Child Protection and SACWIS/CCWIS Technology Professionals.'' 
\kholdedit{Finally, w}\kholddelete{W}e used Prolific to recruit both participants with and without social work backgrounds. To recruit Prolific participants with \kholdedit{background} \kholddelete{domain expertise }in social work, we restricted participation for the study to workers who had completed an undergraduate or graduate (MA/MSc/MPhil) degree in social work. On Prolific, we further restricted participation to workers with a minimum approval rate of 90\%\kholddelete{and who had never taken any of our lab’s prior studies before}.
Participants were compensated \$13 for completing the study. Participants who also had social work domain expertise were compensated an additional \$7 for a total of \$20.

All participants were based in the U.S. \kholddelete{,}\kholdedit{and were} above 18 years of age\kholddelete{, and had never taken the study before}. 
Of \kholdedit{our}\kholddelete{the} 354 participants, 103 participants had domain expertise in social work (i.e., the participant indicated they are currently or were previously a social work graduate student or a practicing social worker). Importantly, only 5 of these 103 participants with social work domain expertise indicated that they have expertise in child welfare specifically, and only 12 participants with social work domain expertise indicated they had prior experience using an AI tool to assist their decisions. However, overall, the 103 participants that have domain expertise in social work may still have more relevant domain knowledge for the training tasks than the general population of workers we recruited from Prolific. We further discuss implications of our recruitment criteria on the interpretation of our findings in the Discussion (Section~\ref{discussion_domainexpertise}).



The study included two comprehension checks and two attention checks that were used to exclude participants from the final dataset. The design of the comprehension and attention checks followed Prolific’s policies \footnote{\url{https://researcher-help.prolific.co/hc/en-gb/articles/360009223553-Prolific-s-Attention-and-Comprehension-Check-Policy}}. 
%
\kholddelete{The two comprehension checks were shown right after being shown brief instructions on how to make their decisions for the AI-Assisted Decision-making activity (in Part 0). The first comprehension check asked: ``Based on the instructions you have just read, what information sources should you consider to make a screening decision in the training activity? There is only one correct answer. Please re-read the instructions above if you are not sure.’’ The second comprehension check asked: ``Based on the instructions you have just read, what is a valid reason for screening out a potential child maltreatment case? Please re-read the instructions above if you are not sure.’’ Both comprehension checks were multiple choice questions with four response options. Participants were given two tries to select the correct choice on each of the two comprehension checks.} 
Participants who failed to answer at least one of the comprehension checks were navigated to the end of the study. 
\kholddelete{The two attention checks were shown during the Training Phase. After participants were shown a set of 8 consecutive cases, they were given an attention check. The attention checks were identical to the other AI-assisted decision-making tasks, except that they included instructions on what response to input.} Participants who \kholdedit{failed at least one}\kholddelete{did not correctly input the instructed response for one or both} of the attention checks were excluded from the final dataset. 

\subsection{Measurement and Analysis} \label{Measurement and Analysis}
\akedit{\khedit{Drawing upon findings from prior field studies  ~\cite[e.g.,][]{cheng2022child,kawakami2022improving},}\khdelete{We sought to measure\khdelete{ changes in} participants' learning with regards to \textit{critical use}: the practice of calibrating reliance on AI predictions, by drawing upon their complementary knowledge and abilities as human expertise.} \khedit{we defined a set of}\khdelete{To capture different aspects of critical use, we defined both} \khedit{\textit{indicators} of critical use in the context of AI-assisted child maltreatment screening, overviewed below.}\khdelete{measures to ground our analysis.}}

\paragraph{Decision alignment measures}


To \khedit{quantitatively study changes in participants' decision-making,}\khdelete{\kholdedit{analyze participants' learning,}} we defined \akdelete{two alignment}\akedit{two \khedit{decision alignment}} measures\khedit{, covering both the ``human'' and the ``AI'' portion of the past recorded AI-assisted decisions used in our study}\khdelete{, informed by past work(e.g., \cite{De-Arteaga2021,kleinberg2018human})}. \kholdedit{
Because our experiment uses real, historical data (as described in Section \ref{case_selection}}), this enables us to explore how participants' decisions compare to past decisions made by workers experienced in AFST-assisted decision-making. 
\akedit{The first of our measures captures agreement with the \textit{AI model}, while the other \khedit{captures agreement with \textit{experienced workers:}}\khdelete{two measures capture different \khedit{notions}\khdelete{aspects} of decision \textit{accuracy}:}}
\begin{itemize}

\item \akedit{\textbf{Model \khedit{Agreement}\khdelete{Alignment}}}: \akedit{We say there are \textit{increases} in Model \khedit{Agreement}\khdelete{Alignment} if 
participants \khedit{come to agree}\khdelete{are more aligned} with the AI prediction \khedit{more often}. \khdelete{For a given case, a participant's decision is \textit{model aligned} if they agreed with the AI prediction and \textit{model misaligned} otherwise.}}
\item \textbf{\akedit{Worker-based Accuracy}\akdelete{Alignment with \kholdedit{past human worker}\kholddelete{the past experienced worker}}}: \akedit{We say there are \textit{increases} in Worker-based Accuracy if participants' decisions \khedit{come to agree with past experienced workers' decisions more often.}\khdelete{are more accurate with respect to} \khdelete{\khedit{the actual AI-assisted decisions made by} past experienced workers' decisions.}
\khdelete{For a given case, a \khedit{screening} decision is \textit{accurate} with respect to the experienced worker's decision if the participant's decision \khdelete{to screen in or screen out }was aligned with \khedit{the worker's.}\khdelete{the past experienced worker's decision to screen in or screen out a given case}.}} 

\end{itemize}

\lgedit{
\akedit{Our \khedit{main analysis examines}\khdelete{primary model analyzes}}\akdelete{Finally, to analyze} participants' learning between the pre- and post-assessments \khedit{with respect to each of these two measures. In particular, we examined interactions between the assessment phase ($Z_{phase}$, a binary indicator with pretest = 0 and posttest = 1) and both participants' prior domain knowledge ($Z_{prior}$) and the presence of explicit feedback during training ($Z_{feedback}$). We also included the participant ID \akedit{$e_{pid}$} as a random effect, to account for the nesting of responses within participants:}
\khdelete{across our alignment and accuracy based measures. We include binary fixed effects term $Z_{phase}$ indicating the assessment phase in addition to $Z_{prior}$ and $Z_{feedback}$ fixed effects terms introduced above. }
\khdelete{We also include $e_{pid}$ as a random effect in these models to account for variation across participants.}  
\khdelete{A positive $Z_{phase}$ coefficient indicates that the dependent variable increased in the post-assessment as compared to the pre-assessment.  We include interaction terms between phase and the other fixed effects, which yields the following model.}
\begin{equation}\label{eq_prepost}
    \hat{Y}_{m} \sim 1 + Z_{phase} * (Z_{prior} + Z_{feedback}) + e_{pid}
\end{equation}

}

\akedit{We supplemented \khedit{our pre-post analyses with analyses of participants' learning across the entire study.}\khdelete{these analyses from Equation~\ref{eq_prepost}}} 
\khedit{In particular, we fit mixed effects models}\khdelete{Equation~\ref{eq_learning} allows us} to examine interactions between the \kholdedit{practice opportunity}\kholddelete{training case} number \akedit{$Z_{num}$} (ranging from 1 - 24) and the presence of explicit feedback \akedit{$Z_{feedback}$} (a binary indicator). 
\akedit{To analyze differences across case categories, we additionally examined interactions between the case type and the practice opportunity number. 
}
As described in Section~\ref{CaseCategorization}, \kholdedit{each case category}\kholddelete{case type (a) - (c) each} applied to 50\% of the 24 total cases, making it possible to run \kholdedit{these}\kholddelete{case type-specific} analyses on balanced amounts of data. \khdelete{\akedit{We again used \akcomment{\hl{n}} variations of Equation~\ref{eq_learning_case} to examine different $\hat{Y}_{m}$ with respect to each of our defined measures.}}

\paragraph{Process-oriented measures} 
\khedit{To \akdelete{measure}study \textit{process-oriented} aspects of critical use, in}\khdelete{In} addition to analyzing our decision alignment measures \khedit{we used Equation~\ref{eq_prepost} to examine pre-post changes in participants' ability to mentally simulate and predict what the AI score would be in particular cases. Participants'}
mean squared error on the AI Score Prediction activity \khedit{is shown as ``Guessing Error'' in Table~\ref{table:pre_post_learning_effects})}.
\khedit{We also}\khdelete{To \akdelete{measure}examine \textit{process-oriented} aspects of critical use, we}\akdelete{both quantitatively and} qualitatively analyzed \khdelete{open text responses (Section \ref{OpenTextResults}).} \akedit{\khdelete{We analyzed two sources of qualitative data: }(1)} \akdelete{we analyzed}participants' open text explanations for how they made their decision\khedit{s}\khdelete{for each AI-assisted decision-making task} in the pre- and post-assessments\akdelete{.}\akedit{, and} 
\khedit{\akdelete{, as well as}\akedit{(2)} their self-reported decision-making goals, collected at the end of the study.}
\khdelete{ of how they made their decisions} \khdelete{Second, we examined open text responses\kholddelete{the open-text responses in the online study.} \akdelete{This included responses from the pre- and post-assessments capturing how participants made their \kholddelete{AI-assisted }decisions, as well as responses} from the end of the \kholdedit{study}\kholddelete{online experiment} capturing their \kholdedit{self-reported decision-making} goals\kholddelete{for making AI-assisted child maltreatment screening decisions}.}

\subsubsection{\akedit{Standard accuracy measure: Proxy-based accuracy}}
The standard measure \khdelete{of accuracy }\khedit{that is used in existing research literature to measure}\khdelete{for} appropriate reliance is accuracy with respect to the \textit{model's predictive target}\khedit{~\cite{guerdan2023ground}}, which we call Proxy-based Accuracy. We say there are \textit{increases} in \textbf{\khedit{Proxy}\khdelete{Model}-based Accuracy} if participants' decisions \khedit{become}\khdelete{are} more accurate with respect to the model's predictive target. For a given case, a decision is \textit{accurate} with respect to the model's predictive target if the participant either screened in a case that resulted in out-of-home placement in two years or screened out a case that did not result in out-of-home placement in two years. 

\subsection{\akedit{Positionality}}
\akedit{Collectively, the authors hold research expertise across human-computer interaction, learning sciences, statistics, machine learning, computer-supported cooperative work, and critical computing studies. The lead author and co-senior authors have research experience observing and designing interventions with child welfare workers who use AI-assisted decision-making tools in their daily work. The study design in this paper reflects both \khdelete{the research team's }knowledge from \khedit{published research}\khdelete{prior} literature and the authors' knowledge gained \khedit{through field studies}\khdelete{from prior empirical research experiences}. None of the authors are affiliated with a child welfare agency; all research was conducted independent of any \khedit{particular child welfare agency.}\khdelete{child welfare agencies and institutions. None of the authors have experience ....} }

\akedit{Importantly, \khedit{we explore}\khdelete{our intention with this paper is to suggest that} \khdelete{setting a higher standard for training---in particular, by }training towards \textit{critical use}\khdelete{---} \khedit{in this paper because we believe doing so} is crucial in many settings where AI systems are deployed. \khedit{As we have argued above, AI performance in complex, real-world settings is bound to be imperfect, and some limitations are to be expected---including some degree of target-construct mismatch and limitations in the information a model is able to access or interpret}. Training towards critical use may mitigate downstream \khedit{negative} impacts of AI model limitations. However, we advise such training only in settings where the deployment of an AI-based decision support tool is actually justifiable in the first place. Training alone \textit{cannot} overcome fundamental design flaws in the AI model or \khedit{surrounding social systems}\khdelete{context of use}.} \khedit{While we examine AI-assisted child maltreatment screening as an example of a highly complex social decision-making setting in this study, we note that existing AI deployments in this domain have seen many fundamental design critiques, including prior research by members of our team ~\cite{cheng2022child,coston2022validity,eubanks2018automating,kawakami2022improving,kawakami2022care,saxena2022unpacking,wang2022against}.}

\section{Results} \label{Results}

\begin{table*}[t]
\begin{center}
\renewcommand{\arraystretch}{1.2}{
\resizebox{\textwidth}{!}{
\begin{tabular}{l *{4}{>{\centering\arraybackslash}m{2.4cm}}}
\toprule
   & \multicolumn{3}{c}{ \textbf{Indicators of Critical Use}} & \textbf{Standard \lgdelete{Accuracy}\lgedit{\khedit{Metric}\khdelete{Indicator}}}  \\ 
\cmidrule(lr){2-4} \cmidrule(lr){5-5} 
   & \textbf{Model 1} & \textbf{Model 2} & \textbf{Model 3} & \textbf{Model 4} \\ 
   & Model Agreement &  Worker-Based Accuracy & Guessing \khedit{Error}\khdelete{Accuracy} &  Proxy-Based Accuracy \\
\hline
\textbf{Fixed Effects}       &            &      &     &    \\        
(Intercept)                  &	 \textbf{0.65}$^{***}$ (0.02)   &   \textbf{0.50}$^{***}$ (0.02) &  \textbf{6.13}$^{***}$ (0.17) &    \textbf{0.73}$^{***}$ (0.02)    \\
Phase \hzedit{(post vs pre)} &    \textbf{-0.41}$^{***}$ (0.03)   & \textbf{0.08}$^{**}$ (0.03) &  \textbf{-3.03}$^{***}$ (0.22)  &  \textbf{\textbf{-0.16}$^{***}$} (0.03)  \\
Prior knowledge              & 	 0.00 (0.03)  &   -0.03 (0.03)       &	 0.08  (0.25)      &    -0.02 (0.03)         \\
Feedback                   	 &   0.04 (0.02)	      &   -0.03 (0.03)        &  -0.26 (0.22)   &    -0.03 (0.03)	    \\
Prior knowledge × Phase	     &   0.02 (0.04)	      &   0.02	(0.04)     &   0.42 (0.31)      &   0.01 (0.04)        \\
Feedback × Phase             &	 0.03 (0.03)         &   0.07 (0.04)   &  \textbf{0.61}$^{*}$ (0.28)	   &  0.05 (0.04)	    \\
\midrule
Observations   & $2832$   & $2832$   &  $2832$  &  $2832$ \\
Marginal $R^2$ & $0.153$  & $0.019$ &  $0.017$  &  $0.105$ \\
\bottomrule
\multicolumn{4}{l}{$^{***}p<0.001$; $^{**}p<0.01$; $^{*}p<0.05$}
\end{tabular}}}
\caption{Coefficient estimates for models examining pre-post learning effects. Each cell shows the coefficient estimate followed by the standard error for a given term (rows) and model (columns). A positive \textit{Phase} coefficient indicates an increase in the measure in the post-assessment as compared to the pre-assessment baseline. A visual summary of these findings is reported in Table~\ref{table:pre_post_learning_effects_visual}.} 
\label{table:pre_post_learning_effects}
\end{center}
\end{table*}

Overall, we found that participants in both conditions started off relying on AI model predictions more \kholdedit{heavily} in earlier practice opportunities. \kholdedit{Early on, p}articipants \kholdedit{aligned their decisions with} the AI model prediction even \kholdedit{on cases} for which the model failed to accurately predict the proxy on which it was trained. However, with \kholdedit{more practice}, participants \khedit{in both conditions} \kholdedit{were more likely to make decisions}\kholddelete{recommendations} that \kholddelete{more likely to disagree}\kholdedit{disagreed} with the AI prediction. 
\khdelete{\khedit{Interestingly, we observed this shift}\khdelete{\kholdedit{Furthermore, we observed this trend}} even \kholdedit{among participants in the \textit{Practice} condition, who}\kholddelete{when they} were not provided with explicit feedback.}

Strikingly, we found that participants learned to \khedit{disagree with AI predictions more often,} \khdelete{make decisions }in ways that aligned more with past frontline workers\khedit{' decisions,} \khedit{even in the \textit{Practice} condition, where participants did not receive} \khdelete{even in the absence of }any explicit feedback about past frontline workers' decisions. An examination of participants' open-text feedback suggests that this occurred because participants learned through practice to integrate case-specific, qualitative details from the allegations when making their decisions, with a focus on assessing nearer-term risks to the child's safety (in contrast to the AI model's focus on predicting \khedit{a proxy outcome, placement,}\khdelete{ placement} on a longer time window). \khedit{Past research suggests that this mirrors the way experienced workers make decisions in the field~\cite{cheng2022child,kawakami2022improving}. In turn,}\khdelete{By focusing more on qualitative case narratives and adopting similar decision-making objectives~\cite{cheng2022child,kawakami2022improving},} participants' decisions began to \khedit{diverge from model predictions and} resemble\mledit{d} those of past experienced workers, with increased practice (Section \ref{OpenTextResults}). 



\kholdedit{\khdelete{We also found that, }Through repeated practice making AI-assisted decisions during the training phase, participants in both conditions} \kholdedit{learned to more accurately predict how the AI model would behave on particular cases.} \akedit{Compared to participants who received both practice and explicit feedback on their decisions,} participants who received \akdelete{\kholdedit{explicit} feedback on their decisions}\akedit{practice alone} saw more improvement in their \kholdedit{ability to predict the AI model's behavior}.\akdelete{ than participants who did not receive feedback.} 

\akdelete{However,} \akedit{Moreover,} receiving explicit feedback on decisions did not \khedit{impact participants' learning with respect to \textit{decision-making}}\khdelete{lead participants to disagree with AI predictions more often}, compared to the effects of repeated practice alone. \khedit{Our analyses suggest that the qualitative case narratives present in both conditions are a rich information source through which participants can learn to calibrate their reliance on AI predictions. Thus, even in the absence of \textit{explicit} feedback on decisions: a training interface that provides accelerated practice, with opportunities to cross-check AI predictions against qualitative narratives, appears to be a stronger baseline than expected.} \khdelete{With more practice, participants' decisions were more likely to align with the actual final decisions made by past experienced workers who were assigned a given case. This was true even for cases in which the past experienced workers’ final decisions did not align with the proxy outcomes that the AI model was trained to predict.}

\khedit{In the following subsections, we first present findings from quantitative analyses of participants' learning and decision-making (Sections \akdelete{4.1-4.4}\akedit{4.1, 4.2, 4.3, 4.5}). In Section 4.4, we present \akedit{complementary} findings from a \akedit{qualitative examination of}\akdelete{mixed-methods analysis, examining changes in} participants' decision-making \textit{processes} across practice opportunities.} Table~\ref{table:pre_post_learning_effects} includes aggregate outcome-based findings, and  Table~\ref{table:pre_post_learning_effects_visual} provides a summary of the directionality of outcome-based findings across the two conditions. 
\khdelete{\akedit{We first present ten findings (Findings 1 - 10) pertaining to the statistical results \hl{Section X}, and then discuss qualitative findings at the end (\hl{Section X}).}}


\subsection{\akedit{Model Agreement:} Participants learn to disagree with AI predictions more often} 
\label{AIAlignmentResults}
\begin{itemize}
\item (Finding-1)  \textbf{Participants disagree with the AI prediction more after repeated practice.}
\item (Finding-2) \textbf{No effect of explicit feedback \khedit{on participants' agreement with AI predictions.}\khdelete{. Participants who received explicit feedback did not learn to disagree with the AI prediction more compared to participants who received repeated practice alone.} }
\end{itemize}

  \begin{figure*}
    \includegraphics[width=\textwidth]{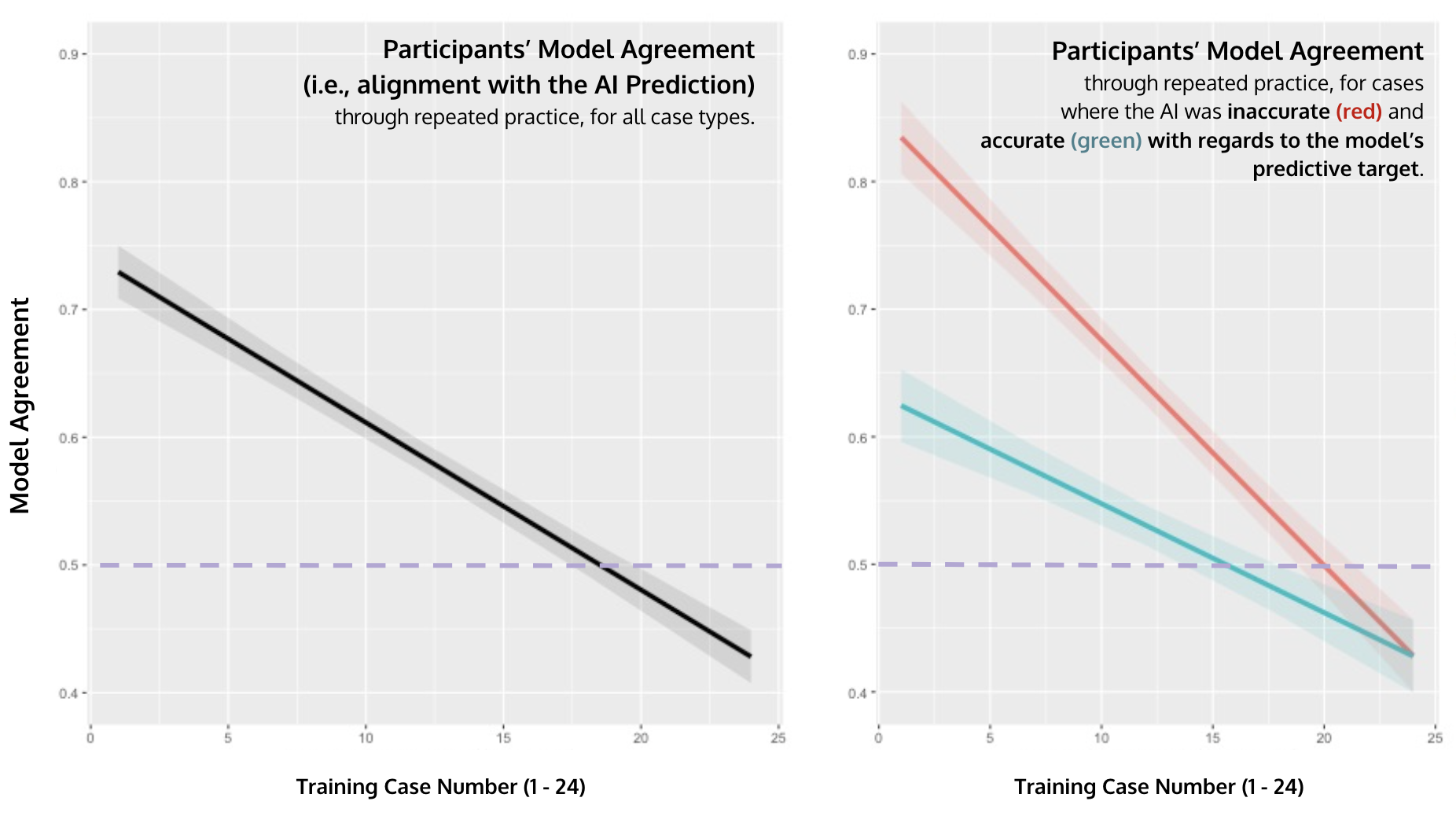}
    \caption{Mixed effects regression visualizing participants' agreement with the AI prediction, across the 24 training cases. Shaded regions indicate standard error. The purple dotted line indicates average accuracy of the AI model with respect to its (proxy) predictive target.}
     \label{fig:afst_align_fig}
\end{figure*} 

Participants started the study by relying on the AI prediction more often, including in cases where the past experienced worker disagreed with the AI prediction. But over time, participants \textit{were less likely to get nudged by the AI prediction}. Instead, over time, participants began to disagree with the AI prediction more often for all case types\hzdelete{, including for cases for which the eventual outcome of the referral suggested the AI prediction accurately predicted its proxy target}.  As shown in Figure~\ref{fig:afst_align_fig}, in the first case, over 70\% of the participants agreed with the AI prediction. However, in the 24th case, only 42\% of the participants agreed with the AI prediction. \akdelete{The study phase coefficient in Table~\ref{table:pre_post_learning_effects}  shows a significant reduction in model agreement in the post-assessment as compared to the pre-assessment (Coef.=-0.407; p$<$.001; 95\% CI: [-.46, -.35]).} \akdelete{Table~\ref{table:pre_post_learning_effects} shows a 32\% decrease in the \akdelete{AI agreement}\akedit{Model \khedit{Agreement}\khdelete{Alignment} measure}.} 
This finding was corroborated by the learning effects regression (Table~\ref{table:pre_post_learning_effects}), which showed a significant decrease in Model \khedit{Agreement}\khdelete{Alignment} between pre- and post-assessments \lgedit{as indicated by the study phase coefficient \lgdelete{(Coef.=-0.315; p<.0001; 95\% CI: [-.35, -.29])} \lgedit{(Coef.=-0.407; p$<$.001; 95\% CI: [-.46, -.35])}.}

\subsection{Worker-based Accuracy: Participants become \textit{more} aligned with experienced workers} \label{WorkerAlignmentResults}
 \begin{figure*}
    \includegraphics[width=\textwidth]{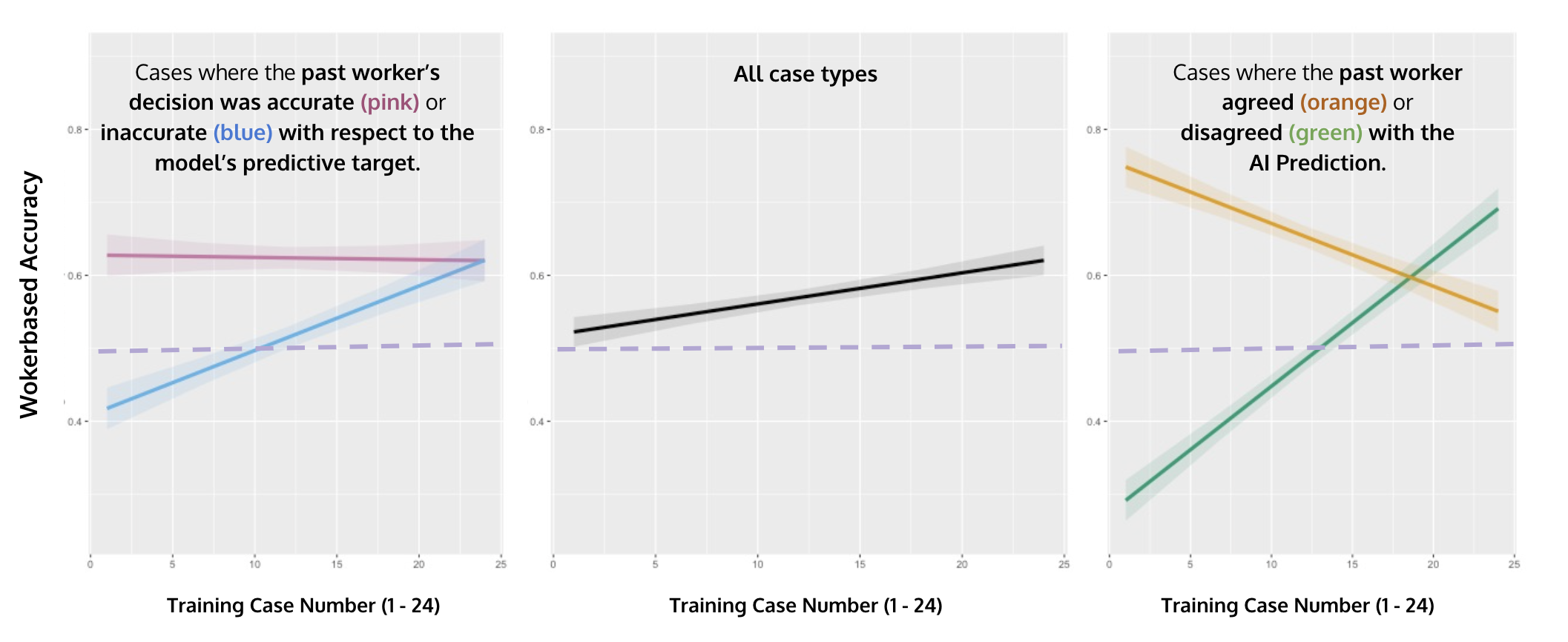}
    \caption{Mixed effects regression visualizing participants' learning across repeated practice, with respect to the worker-based accuracy measures (i.e., accuracy with respect to past experienced workers).}
     \label{fig:worker_align_fig}
\end{figure*} 
\begin{itemize}
    \item (Finding-\akedit{3}\khdelete{5,7}) \textbf{Participants’ decisions \akedit{\khedit{become}\khdelete{are} more accurate with respect to}\akdelete{align more with}\khdelete{ the past} experienced worker\khedit{s'} decisions after repeated practice. }
    \item (Finding-\akedit{4}\khdelete{6,8}) \akdelete{\textbf{Participants who received explicit feedback \akedit{had more accurate decisions with respect to the}\akdelete{aligned more with the} past experienced worker, compared to the participants who received repeated practice alone.}
    \khcomment{Should Finding 8 be changed to say that there was no effect of feedback (given updated results in Table 1)?}} \akedit{\akdelete{\textbf{Participants who receive explicit feedback have less accurate decisions with respect to the AI model's predictive target, compared to the participants who received repeated practice alone.} \khcomment{It seems like Finding 4 should be updated to say that there is no effect, given the updated results in Table 1?}}
    \textbf{No effect of explicit feedback \khedit{on the accuracy of participants' decisions with respect to experienced workers' decisions.}\khdelete{. Participants who received explicit feedback did not learn to have more accurate decisions with respect to the past experienced worker decisions, compared to participants who received repeated practice alone.}} }
\end{itemize}

\kholdedit{With practice,} participants’ \kholdedit{decisions began to align more with the final decisions of past experienced workers}. Overall, Figure~\ref{fig:worker_align_fig} (middle) shows that, at the start of the training, 52\% of the participants made decisions that \akedit{were accurate with respect to the past experienced worker.}\akdelete{aligned with the past experienced worker.} By the end of the training, 61\% of the participants made decisions that \akedit{were accurate with respect to the past experienced worker.}\akdelete{aligned with the past experienced worker. } This finding is consistent with learning effects regression results, which show a significant increase in Worker-based Accuracy scores between pre- and post-assessments \lgdelete{(Coef.=0.157; p<.0001; 95\% CI: [.13, .18])} \lgedit{(Coef.=0.083; p$<$.01; 95\% CI: [0.027, 0.140])}.

With increased practice, participants\akedit{' decisions} were especially likely to \akedit{be accurate with respect to}\akdelete{align with} the past experienced worker when \textit{disagreeing }with the AI prediction. As shown in Figure~\ref{fig:worker_align_fig}, for the category of cases in which the past experienced worker disagreed with the \akedit{AI prediction}\akdelete{AFST score} (right), for the first case, only 30\% of participants made decisions that \akedit{were accurate with respect to the past experienced worker}\akdelete{aligned with the past experienced worker} and \akedit{in disagreement}\akdelete{misaligned} with the AI prediction. By the 24th case, nearly 70\% of participants made decisions that \akedit{were accurate with respect to the past experienced worker}\akdelete{aligned with the past experienced worker} and \akedit{in disagreement with}\akdelete{misaligned with} the AI prediction. On the other hand, for the category of cases in which the past experienced worker agreed with the \akedit{AI prediction}\akdelete{AFST score}, at the start of the training, nearly 75\% of participants made decisions that \akedit{were accurate with respect to}\akdelete{aligned with} the past experienced worker. By the 24th case, this proportion went down to just over 50\% of participants.  

Additionally, participants\akedit{' decisions became more accurate with respect to the past experienced worker} \akdelete{began to align more with past experienced workers} regardless of whether those past experienced workers made decisions that were \akedit{inaccurate or accurate with respect to the model's predictive target}\akdelete{outcome aligned}\akdelete{(i.e., a screen in decision leads to placement out of home in two years) or misaligned}, a\akedit{n accuracy-based} measure that is widely used in prior studies evaluating the accuracy of AI-assisted decision-making with \akedit{this AI model, the} AFST (e.g., in ~\cite{cheng2022child,De-Arteaga2020}). As shown in Figure~\ref{fig:worker_align_fig}, for cases in which the past experienced worker was \akedit{inaccurate with respect to the model's predictive target}\akdelete{outcome aligned} (left), around 60\% of participants at both the start and end of the training made decisions that \akedit{were accurate with respect to}\akdelete{aligned with} the past experienced worker. However, for cases in which the past experienced worker was \akedit{accurate with respect to the model's predictive target}\akdelete{outcome misaligned}, only around 40\% of participants made decisions that \akedit{were accurate with respect to}\akdelete{aligned with} the past experienced worker in the first case. Yet, by the 24th case, this number increased to around 60\% of cases, erasing any differences in decision behavior for cases in which past experienced workers were \akedit{accurate versus inaccurate with respect to the model's predictive target}\akdelete{outcome aligned versus misaligned}. \akdelete{While these patterns of \akedit{making more accurate decisions with respect to}\akdelete{aligning more with} the past experienced worker through the training were observed for all participants, participants who received feedback on their decisions throughout the training made decisions that \akedit{were even \textit{more} accurate with respect to the}\akdelete{aligned\textit{ more }with} past experienced workers compared to participants who did not receive feedback. }As shown in Table \ref{table:pre_post_learning_effects}, \akedit{explicit feedback did not have}\akdelete{feedback has } a significant effect \akedit{on Worker-based Accuracy}.

%
\khdelete{However, interestingly, participants indicated they would update their \kholdedit{decision} after seeing the feedback for only ~20\% of the cases. This included 156 out of 177 participants (in the \textit{Practice + Explicit Feedback} condition), with the majority (~70\%) indicating they would update their \kholdedit{decision} for 1 to 3 cases.}

\renewcommand{\arraystretch}{2.0}
\begin{table*}[t]
\begin{tabular}{p{5.5cm} p{2.6cm} p{2.6cm} p{2.6cm} p{2.6cm}}
\toprule
   & \multicolumn{3}{c}{ \textbf{Indicators of Critical Use}} & \textbf{Standard Metric} \\ 
\cmidrule(lr){2-4} \cmidrule(lr){5-5} 
   & \textbf{Model 1} \newline Model \newline Agreement & \textbf{Model 2}  \newline Worker-Based \newline Accuracy & \textbf{Model 3}  \newline Guessing \newline \khedit{Error}\khdelete{Accuracy} & \textbf{Model 4} \newline Proxy-Based \newline Accuracy \\
\hline
\textbf{\textit{Practice Condition} }     & \cellcolor{blue!25}$\downarrow$   &  \cellcolor{purple!40}$\uparrow$  &   \cellcolor{blue!60}$\downarrow$ $\downarrow$ &  \cellcolor{blue!25}$\downarrow$   \\      
\midrule
\textbf{\textit{Practice + Explicit Feedback Condition} }   & \cellcolor{blue!25}$\downarrow$   &  \cellcolor{purple!40}$\uparrow$     &   \cellcolor{blue!25}$\downarrow$  &  \cellcolor{blue!25}$\downarrow$     \\   
\bottomrule
\end{tabular}

\caption{Visual summary of model findings reported in Table~\ref{table:pre_post_learning_effects}. Upwards arrow (rose-colored cell) indicates increases in the measure value after the training. Downwards arrow (blue cell) indicates decreases in the measure value after the training, and two downwards arrows (darker blue cell) indicates greater decreases in the measure value in the repeated practice condition compared to when explicit feedback was shown alongside repeated practice.\akdelete{ \textit{Diversity of keywords} indicates the number of information sources recalled in open text explanations of AI-assisted decisions, in the post-assessment compared to the pre-assessment. An upwards arrow indicates increased diversity of keywords recalled.}} 
\label{table:pre_post_learning_effects_visual}
\end{table*}

\subsection{\akedit{Guessing Error: Participants learn to make improved predictions of the AI model's behavior}}
\label{GuessResults}
\begin{itemize}
\item (Finding-5) \textbf{Participants \khedit{improved in their ability to predict the AI model's behavior}\khdelete{got better at predicting the AI prediction} \khedit{after repeated practice making AI-assisted decisions}\khdelete{after repeated practice}.} 
\item (Finding-6) \textbf{\khedit{Explicit feedback led to \akdelete{greater}\akedit{lesser} improvement in participants' ability to predict the AI model's behavior\khdelete{prediction performance}, compared to practice alone.}}
\end{itemize}

While disagreeing with the AI prediction more often with time, participants were simultaneously \textit{getting better at guessing \khedit{what} the \khdelete{AI predictions}\khedit{AI prediction would be on a given case}}. Participants who received \akedit{practice alone, compared to those who additionally received feedback on their decisions,} \akdelete{alongside practice }saw somewhat more improvement in their guessing performance\akdelete{ than participants who did not receive feedback on their decisions}. As shown in Table \ref{table:pre_post_learning_effects},  there was a significant decrease in guess mean squared error in the post-assessment as compared to the pre-assessment \lgedit{(Coef.=-3.03; p$<$.001; 95\% CI: [-3.45, -2.60])}. \lgdelete{(Coef. -2.312, p $<$ .0001, 95\% CI: [-2.523, -2.101])} Additionally, participants assigned to the \textit{Practice\akdelete{ + Explicit Feedback}} condition saw a greater improvement in their guessing performance as compared to participants in the \textit{Practice\akedit{ + Explicit Feedback}} condition \lgedit{(Coef.=0.608; p$<$.05; 95\% CI: [0.05, 1.15])}. \lgdelete{(Coef. 0.608, p $<$ 0.05, 95\% CI: [.34, .88])} \akedit{Overall,} this result indicates that, despite disagreeing with predictions, participants \akedit{in both conditions} were actively engaging with the learning activity and improving their skills at mentally simulating the behavior of the model. \akedit{However, participants who did \textit{not} receive explicit feedback on their AI-assisted decisions improved these skills more. As we discuss in Section~\ref{discussion_feedback}, it is possible that, compared to the explicit feedback, the information reported in the case (including the qualitative \khedit{case narratives}\khdelete{allegations}) served as a more powerful \khedit{signal through which to learn how the AI model behaves on different cases, compared to the explicit feedback signals provided. It is possible that the \textit{Practice} condition saw greater improvement because explicit feedback on AI decisions distracted participants from learning to predict model behavior based on qualitative narratives.}\khdelete{learning signal to participants than expected.}}
\lgolddelete{<guessing>, X\% of participants saw improvements in their ability to guess the AI prediction in the pre-assessment compared to the post-assessment. There was a 2.41 point decrease in the gap between the participants’ guesses and the actual AI prediction, in the post-assessment compared to the pre-assessment. Participants who received feedback were more likely to see improvement in their guessing ($p = 0.0237 < 0.05$). [Having domain expertise did not predict improvement in guessing ($p = 0.376$).]}

\subsection{Process Analyses: Participants learn to make decisions more like experienced workers}
\label{OpenTextResults}

\khdelete{Put together, Finding-2 and Finding-8 
suggest that participants \textit{selectively learned} from a subset of the six proxies shown as feedback on their decisions. In particular, the findings provide evidence that participants \textit{at least learn from feedback on the supervisors’ decisions} but \textit{not the AI prediction’s observed outcome} (i.e., placement out of home). However, it appears that participants are not necessarily learning from the explicit feedback alone.}
\khdelete{Finding-1 and Finding-7 together also suggest that, through \textit{repeated practice}, participants may have learned to make decisions like the past experienced worker using information they found meaningful in the \textit{case referral}. Participants were learning to disagree with the AI prediction and align with the past experienced worker, \textit{even when} they never saw feedback showing the past experienced worker's decision.}
\akedit{
To understand participants' decision-making process, we qualitatively examined their open text responses\khedit{. These included (1) participants responses} \khedit{in each of the pre- and post-assessments,} where they explained \khdelete{(1) }how they made their AI-assisted decisions\khdelete{, which we asked for each case in the pre-assessment and post-assessment phases of the study} and (2) \khedit{participants' explanations at the end of the study, describing} what \khdelete{decision-making }goals guided their decisions.\khdelete{AI-assisted decisions, which we asked at the end of the experiment.} 

\begin{itemize}
    \item (Finding-7) \textbf{Participants' explanations for their decisions reference their use of model unobservables to help them make AI-assisted decisions.}
    \item (Finding-8) \textbf{Participants' explanations of their decision-making goals indicate they target ensuring immediate safety to the child when making AI-assisted decisions.}
    \khdelete{\item \akedit{(Finding-\khedit{8}\khdelete{10}) \textbf{Participants who received explicit feedback \khedit{gave explanations for}\khdelete{had explanations for} their decisions \khedit{that referenced}\khdelete{recalled that} more information sources, including \khedit{model unobservables.}\khdelete{more allegation-specific information, compared to the participants who recived repeated practice alone.}}}}
\end{itemize}
}
\akdelete{
\begin{itemize}
    \item \akedit{(Finding-\khedit{7}\khdelete{9}) \textbf{Participants' explanations for their decisions \khedit{reference}\khdelete{recall} more information sources, including \khedit{model unobservables}\khdelete{more allegation-specific information}, after repeated practice.}}
    \khdelete{\item \akedit{(Finding-\khedit{8}\khdelete{10}) \textbf{Participants who received explicit feedback \khedit{gave explanations for}\khdelete{had explanations for} their decisions \khedit{that referenced}\khdelete{recalled that} more information sources, including \khedit{model unobservables.}\khdelete{more allegation-specific information, compared to the participants who recived repeated practice alone.}}}}
\end{itemize}
}

\subsubsection{\akedit{Informing decisions using qualitative case narratives.}}
\akdelete{With practice, participants' explanations of how they made their AI-assisted decisions included a larger number of information sources than at the start of the training. As shown in Table~\ref{table:pre_post_learning_effects}, the learning effects regression on the pre- and post-assessments show a significant increase in keyword diversity \akedit{(which includes keywords pertaining to the allegation information shown in the case referrals, amongst other information}) in the Post-assessment explanations. \akdelete{Participants in the \textit{Practice + Explicit Feedback} condition used a larger diversity of keywords compared to participants in the \textit{Practice} condition. Moreover, of the different information sources available to them, participants \mldelete{recalled}\mledit{referenced} allegation-specific information sources more often after practice. Table~\ref{table:pre_post_learning_effects} shows that the number of allegation-specific keyword \mldelete{recollections}\mledit{references} significantly increased in the post-assessment compared to the pre-assessment.} \akedit{There were no explicit feedback or domain expertise effects.}\akdelete{Participants who in the \textit{Practice + Explicit Feedback} condition used a greater number of allegation-specific keywords in their explanations compared to participants in the \textit{Practice} condition. Additionally, participants who had \khedit{prior knowledge related to}\khdelete{domain expertise in} social work also used more information sources, and more allegation-specific information sources, compared to participants who did not have domain expertise in social work.} }
\akedit{An examination of} participants' explanations of how they made their AI-assisted decisions in the pre-assessment and post-assessment phases of the study \akedit{provide \akdelete{additional }evidence of how they relied}\akdelete{suggest that they relied heavily} on the qualitative \khedit{case narratives}\khdelete{allegations} to inform their decisions. Much like how actual experienced social workers used \akdelete{ADS}\akedit{AI-based decision support} tools in their day-to-day jobs, participants drew on context-specific details from the allegations to inform their decisions. For example, participants sometimes drew on their interpretations of the qualitative \khedit{case narratives}\khdelete{allegations}, to inform decisions to disagree with the AI prediction. When explaining their decision to screen out a case with an AI prediction of 16, one participant described: 
``Despite the high risk score[,] I do not see evidence of maltreatment based on the details. The parent was angry at a third party and may have roughly grabbed the child but that does not rise to maltreatment unless it can be proven that there is a pattern of rough handling.’’ 

Similarly, another participant considered missing information in the allegation (e.g., about intentions and causes) to appropriately weigh the severity of the reported allegation, in comparison to the AI prediction: ``Although the risk score is 12, the child appears to be physically healthy and has a positive mood. There are many reasons a knife could be in a bedroom, and it is possible the kid's struggles socially and academically could be caused by a learning disability.’’ In another case,  a participant explained a decision to screen in a case, despite a low AI prediction, given allegations reporting violent behavior: 
``Risk score appears too low for this case.  Multiple examples of violence brought on by behavioral problems of the mother in the home, as well as child acting out at school, all meet definition of legal maltreatment.’’

In other cases, participants referenced the allegations to gain more insight into the child's current living situation, and whether there are signals indicating there may be an upwards or downwards trend for improvement. For example, in explaining their decision to screen out a case, the participant described how a family appears to be on a path towards recovering from past challenges: ``The mother has a support system in place. She has the grandmother, and is receiving public benefits/services. Her treatment team reports that she should be able to fully recover and care for her child. And the child is currently clean and well cared for. '' In another case, a participant noted the lack of support and resources that the child currently has, alongside the history of substantiated referrals, to inform their decision to screen in a case:  ``The fact that the child is not taking her prescribed medication, not receiving counseling for her mental condition, and at risk of homelessness demonstrates that the child is in severe risk of physical and emotional harm. Moreover, the substantiation of past referrals lends further credibility and urgency to the current one.''

Overall, participants did not mention the AI prediction as often when explaining their decisions. In particular, participants mentioned using a ``score'' (i.e., the "AI Risk Score'' shown in the case referral) in 8.72\% of the explanations (for 247 out of 2832 total cases). Explanations considering information from a ``score’’ decreased after the Training Activity Phase, corroborating findings that participants relied on the AI prediction less with increased practice. In the pre-assessment Phase (cases 1 - 4), participants considered the ``score'' for 194 cases (79\% of the 247 mentions). In the post-assessment Phase (cases 21 - 24), the frequency dropped to only 57 cases (21\% of 247 mentions). In many of these cases, participants referenced the score as one of multiple information sources they drew on to explain their decision (e.g., ``The Risk score is a 20, there is commotion and drug use within the family.''). However, in 17 of the 57 cases that mentioned a ``score'' in the post-assessment Phase, participants were describing why they were disagreeing with it (e.g., ``even if the risk score is fairly low, the situation doesn't seem to be very under control and could break any moment''). The 247 mentions of a ``score'' across the pre- and post-assessments were from 112 total participants, split roughly evenly across the \textit{Practice} and \textit{Practice + Explicit Feedback} conditions. 
\subsubsection{\akedit{Making decisions with the goal of ensuring immediate safety.}}
Beyond leveraging their unique ability to access and integrate context-specific information when making decisions, the goals that participants had when making decisions may also explain the learning effects. In particular, it is possible that participants learned to disagree with the AI prediction, while learning to make decisions like past experienced workers, partly because their own decision-making goals aligned more strongly with those of the past experienced workers. At the end of the experiment, participants were asked to describe their own goals when making child maltreatment screening decisions. Many participants described the importance of considering near-term safety risks or harms to the child, which is misaligned with the longer-term outcomes that the AI model is trained to predict. Both participants with and without domain knowledge in social work described goals related to ``immediate safety and harm,’’ aligning with the goals documented of the actual past experienced social workers in this domain. For example, one social worker described: ``I am looking for immediate safety concerns first. Then making sure physical needs are being met. If either of these are a concern, it is an immediate screen in.’’ Another social work graduate student described the importance of ``current’’ evidence:  ``I assess first for evidence of maltreatment or harm currently going on in the home and then consider if additional investigation is necessary to rule out possible abuse or neglect, I then consider if there is enough "probable cause" to truly investigate this further.’’ A participant without social work domain expertise similarly described: ``My goals were whether or not there would be any immediate danger posed to the child either through their environment or by their parents.’’

\begin{figure}
    \includegraphics[scale=0.40]{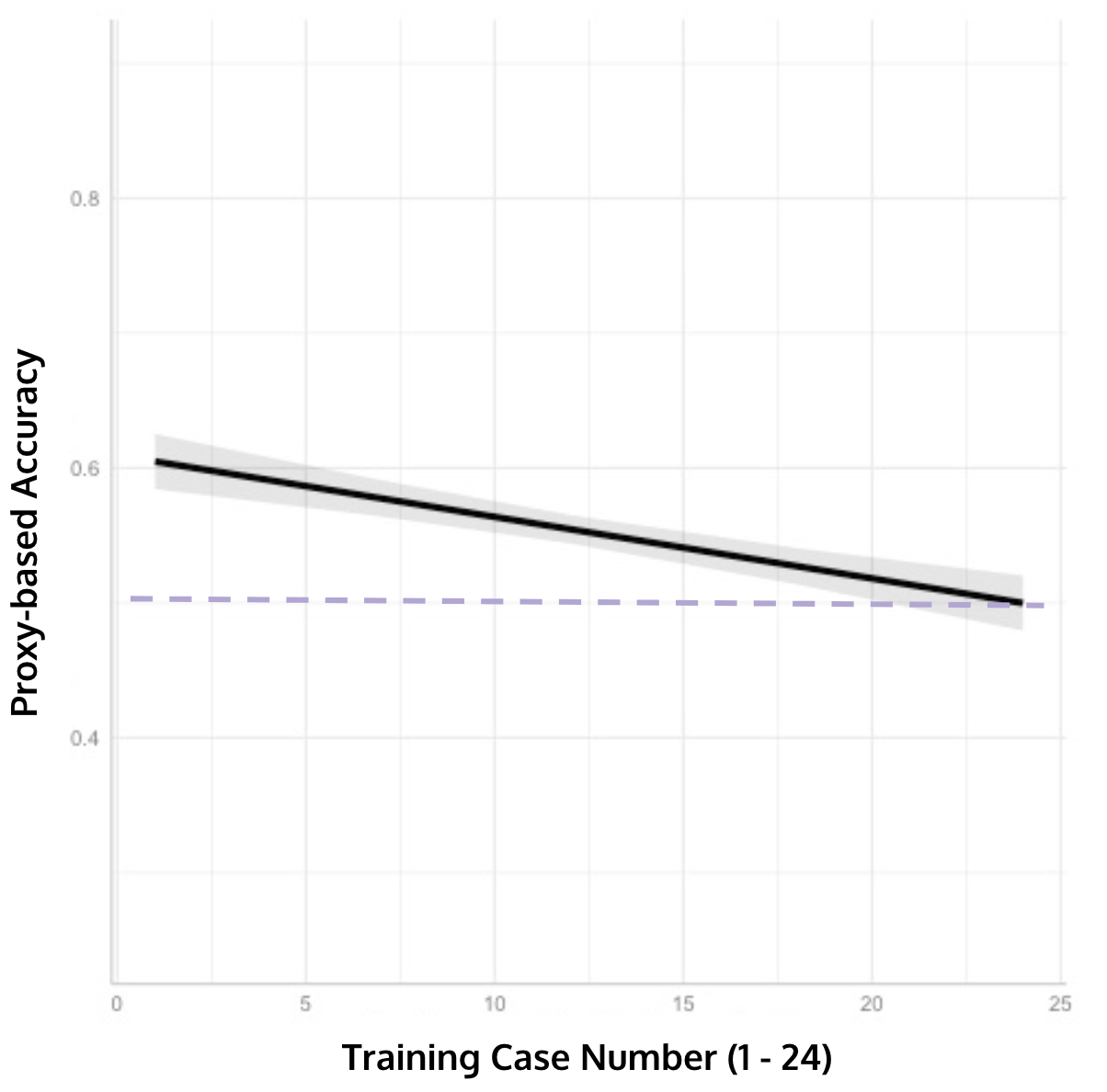}
    \caption{Participants' Proxy-based Accuracy measures against the baseline accuracy of the model (in purple).}
     \label{fig:model_accuracy_fig}
\end{figure}

\subsection{Proxy-based Accuracy: Participants become \textit{less} aligned on the model's targeted proxy} \label{ProxyBasedAccuracyResults}

\begin{itemize}
    \item \akedit{(Finding-\khedit{9}\khdelete{3}) \textbf{Participants’ decisions \khedit{become}\khdelete{are} less accurate with respect to the \khedit{proxy outcome targeted by the} AI model\khdelete{'s predictive target} after repeated practice.} }
    \item \akedit{(Finding-\khedit{10}\khdelete{4}) \akdelete{\textbf{Participants who receive explicit feedback have less accurate decisions with respect to the AI model's predictive target, compared to the participants who received repeated practice alone.} \khcomment{It seems like Finding 4 should be updated to say that there is no effect, given the updated results in Table 1?}}
    \textbf{No effect of explicit feedback \khedit{on the accuracy of participants' decisions with respect to the proxy targeted by the AI model.}\khdelete{. Participants who received explicit feedback did not learn to have less accurate decisions with respect to the proxy outcome targeted by the AI model, compared to participants who received repeated practice alone.}} }
\end{itemize}

Participants gradually began to disagree with the AI prediction regardless of whether the AI was \akedit{inaccurate or accurate with respect to the model's predictive target}\akdelete{outcome misaligned}. As shown in Figure~\ref{fig:afst_align_fig}, for the first case in which the AI was \akdelete{outcome aligned}\akedit{accurate with respect to the model's predictive target}, 80\% of participants agreed with the AI prediction. For the first case in which the AI was \akedit{inaccurate with respect to the model's predictive target}\akdelete{outcome misaligned}, 60\% of participants agreed with the AI prediction. However, in the 24th case, only 40\% of participants agreed with the AI prediction, regardless of whether the \akedit{AI} prediction was \akedit{accurate or inaccurate with respect to the model's predictive target}\akdelete{eventually outcome aligned or misaligned}. 

\kholdedit{Importantly, t}\kholddelete{T}his gradual disagreement with the AI prediction occurred \textit{regardless of whether the participant received feedback on their decisions}. As shown in Table~\ref{table:pre_post_learning_effects}, there is no feedback effect \lgolddelete{ (\akcomment{replace this if using single table}$p = 0.479 > 0.05$ \lgoldedit{This result can be replaced with the statistic at the end of the sentence})} on participants’ decisions to agree or disagree with the AI prediction over time \lgdelete{(Coef. 0.022, p < .50, 95\% CI: [-.01, .05])} \lgedit{(Coef. 0.02, p $<$ .50, 95\% CI: [-0.03, 0.09])}.

\lgcomment{In the section above, I wonder if it could be worth spending a bit of space discussing results on the Proxy-Based Accuracy. The subsection title and findings reference the Proxy-Based Accuracy while the text primarily discusses Model Agreement. Some discussion of Proxy-Based Accuracy findings could address reviewer concerns that this piece was missing from our earlier submission. \sccomment{agree}} \khcomment{Agreed, just a sentence or two directly linking to Findings 9 and 10 could do the trick.}
  
\section{Discussion} \label{Discussion}

In this paper, \khedit{we define}\khdelete{By defining} a \khedit{notion}\khdelete{concept} 
of \khedit{appropriate reliance called} \textit{critical use}, which emphasizes human decision-makers' ability to situate AI predictions against knowledge that is uniquely available to them but unavailable to the AI model. Through an experimental investigation of how training can support critical use of AI predictions in making AI-assisted child maltreatment screening decisions, we explore the effects of repeated feedback and explicit feedback on \textit{what} and \textit{how} participants learn. In the following, we discuss interpretations of our results and avenues for future work. 

\subsection{Evidence of \textit{critical use} from training}
Our findings suggest that \khedit{training}\khdelete{the training experiment} improved \textit{critical use} of the AI tool---humans' ability to situate AI predictions against potentially complementary knowledge available to them (but not the AI model). \akedit{\khedit{A}\khdelete{An extensive} body of prior \khedit{research on}\khdelete{literature in} AI-assisted child maltreatment screening has documented how social workers draw on their knowledge of qualitative, contextual information to \khedit{calibrate their reliance upon AI-based decision support}\khdelete{disagree with the AI tool} (e.g.,~\cite{,kawakami2022improving,Saxena2021,saxena2022unpacking}). \khedit{In doing so, experienced workers can overcome some of the challenges posed by target-construct mismatch and limitations in the information available to AI models, mitigating} erroneous AI outcomes and reducing disparities in decisions \cite{cheng2022child,De-Arteaga2020}.} \akcomment{maybe include more citations here} 
\khedit{In our study, we found that \textbf{participants learned to calibrate their reliance upon AI predictions 
in ways that resembled experienced workers.}}
%
%
In particular, with increased practice, participants learned to disagree more often with AI predictions (Section~\ref{AIAlignmentResults}) \akedit{and, in turn,}
\akdelete{\khedit{while drawing upon a wider range of}\khdelete{using more} information sources \khedit{to inform their decisions}\khdelete{to help explain their decisions} (Section~\ref{OpenTextResults}).\khedit{In turn,}}\khdelete{Simultaneously,}\akdelete{participants}
\khedit{came to make}\khdelete{made} decisions that aligned \khedit{more} with \khedit{those of} past workers with extensive experience making AI-assisted decisions (Section~\ref{WorkerAlignmentResults}). \akedit{Moreover, \khedit{through practice making AI-assisted decisions} participants improved their ability to \khedit{mentally simulate the AI model and predict how it would behave on specific cases}\khdelete{simulate the AI model's behavior}\khdelete{after receiving practice opportunities} (Section~\ref{GuessResults}).} \akedit{\khdelete{Our process-oriented analyses suggest reasons why these patterns of disagreement may have occurred: }Our qualitative examination of participants' explanations suggest that participants drew upon \khdelete{their }knowledge regarding the actual decision-making goals (i.e., screening for short\khedit{er-}term safety) and interpreted qualitative \khedit{narratives unavailable to the AI model in order}\khdelete{, context-specific information sources (i.e., allegation information)} to inform their AI-assisted decisions (Section~\ref{OpenTextResults}) and learn how the AI model behaves on different kinds of cases.}\khdelete{\akedit{These qualitative narratives may also have also helped participants hone their ability to predict the AI model's behavior.}} 
\khdelete{\akedit{Past research indicates that this process---leveraging qualitative information to inform when to disagree with AI predictions---is a powerful skill that enables social workers to \khedit{rely more appropriately upon imperfect AI predictions, in the presence of target-construct misalignment and information asymmetries~\cite{cheng2022child,coston2022validity,kawakami2022improving}.}}}
%
%
\akedit{However, \khedit{we note that future work is needed to understand the \textit{extent} to which specific downstream impacts of target-construct mismatch and information asymmetries (e.g., systematic errors and unfairness in decision-making) can be mitigated through training.}\khdelete{our claim opens questions for future work and reflection: we cannot claim through our study that \textit{critical use} \textit{overcomes} (rather than helps mitigate) these challenges, nor does our study make clear \textit{to what extent} these challenges were mitigated through training.}} 

Analyses using a standard measure of accuracy for human-AI decision-making would indicate that increased practice led participants to make \textit{less} accurate decisions, when accuracy is measured with respect to the model's target proxy (Section~\ref{ProxyBasedAccuracyResults}). Interpreting \khedit{proxy-based accuracy}\khdelete{these proxy-based accuracy findings} in isolation from the rest of our analyses \khedit{may lead to the conclusion}\khdelete{ suggest } that participants \khedit{are}\khdelete{may} simply \khdelete{be }falling into patterns of algorithm aversion~\cite{burton2020systematic,dietvorst2015algorithm}. 
\akedit{However, in analyzing \khdelete{process- and outcome-oriented }indicators of \textit{critical use}, \khedit{including both process- and outcome-oriented \akdelete{measures}\akedit{signals},} our findings \khedit{suggest that participants disagreed with AI outputs in sophisticated ways.} 
\khcomment{Trimming some text below to reduce redundancy with points that have already been made above.} 
\khdelete{shed light onto a different narrative: rather than simply disagreeing with AI outputs due to, for example, a distrust in the AI model, participants were disagreeing in ways that mimicked past experienced social workers by leveraging information sources that only they, as human decision-makers, had access to.}} Indeed, in complex, real-world decision-making settings, common outcome-focused metrics for evaluating AI-assisted decisions---for example, comparing human decisions with the AI model's ground truth label---can provide an incomplete understanding of the quality of each decision~\cite{kawakami2022improving,guerdan2023ground}. 
\akedit{\khedit{Our findings point to a}\khdelete{There is a} compelling opportunity \khedit{to improve}\khdelete{for improving} evaluations of AI-assisted decision-making in other complex domains, by designing \khedit{and employing context-specific measures of \textit{critical use, in addition to examining proxy-based accuracy}.}\khdelete{measures on a context-specific understanding of \textit{critical use} indicators in a given setting.} Moreover, to expand the set of indicators that could be used to measure decision-making, future work should innovate on ways to better capture process-based signals for decision quality (e.g.,~\cite{kyeremanteng2015process,mcauliffe1979measuring}). We further discuss how our study findings and approach may generalize across domains in Subsection~\ref{discussion_generalize}. }

\paragraph{Limitations}
For the purposes of our study, we selected cases where the AI tool had an average proxy-based accuracy of (50\%). This is lower than the the AFST model's actual accuracy with respect to its proxy. Our decision to include cases that aggregated to 50\% proxy-based accuracy of the AI tool was informed by our goal of having a well-balanced set of examples to conduct case type-specific analyses and of prior literature discussing the risks of over-reliance in the absence of sufficient examples of the AI model erring~\cite{passi2022overreliance}. 
While we cannot rule out that this contributed to increased disagreement with the AI tool, our findings indicate that it is \textit{not sufficient} to explain participants' learning. For example, participants aligned less with the model proxy following training even in the \textit{Practice} condition, when participants did not receive explicit feedback on the accuracy of the AFST with respect to its targeted proxy (see ~\cite{lu2021human}). 

\subsection{Why did explicit feedback on AI-assisted decisions not enhance learning?}\label{discussion_feedback}
We \khedit{did not observe significant impacts of}\khdelete{found that} showing explicit feedback on \khdelete{\khedit{our \akedit{quantitative} indicators of critical use. 
\akedit{Explicit feedback} \akdelete{it }did not lead to significant changes in }} participants' \textit{decision-making}, compared with practice alone. 
\khdelete{decisions did not lead to improved decisions along the critical use indicators.}
Furthermore, although \akedit{the training improved participants' ability to predict how the AI model would behave on a given case,} participants who received practice opportunities \textit{without} explicit feedback saw greater improvements than those who received feedback. 
\khedit{Taken together, our}\khdelete{These} findings suggest that \khedit{when it comes to learning to make AI-assisted decisions,} the \khedit{qualitative case narratives}\khdelete{case-specific information} participants see and interpret (which are inaccessible to the AI model) may have had a bigger impact than we had originally anticipated. These narratives, which were present in both conditions, served as a rich information source \khedit{regarding the plausibility of individual AI predictions,} through which participants were able to learn to calibrate their reliance on AI predictions. Thus, even in the absence of \textit{explicit} feedback on decisions, a training interface that provides accelerated practice, with opportunities to cross-check AI predictions against qualitative narratives---a form of \textit{implicit} feedback on the reliability of individual AI predictions---appears to be a stronger baseline condition than expected. \akedit{\khedit{Our findings further suggest}\khdelete{This further suggests} that participants who were shown explicit feedback \khedit{may not have perceived}\khdelete{did not recognize} the feedback \khedit{provided}\khdelete{(i.e., the selection of observed outcomes)} as useful signals of decision quality\khedit{, relative to the signals provided through the qualitative case narratives}}.

\khcomment{Trimmed the previous first sentence of this paragraph to reduce redundancy with points already made a few sentences before. I think we can just dive into the next sentence about prior literature.}
\khdelete{Moreover, in our study, the qualitative case narratives served as a rich source of information to help participants calibrate their reliance on AI predictions.} 
\akedit{\khedit{Prior work}\khdelete{Prior theories} from the learning sciences \khedit{suggests that} for certain learning tasks, providing \khedit{learners}\khdelete{students} with ``grounded'' feedback---\khedit{concrete representations that offer rich,}\khdelete{accessible feedback representations that provides} meaningful signals \khedit{about whether or not an individuals' targeted outcome is achieved}\khdelete{to an individual's desired target outcome}---\khedit{is}\khdelete{has proven to be} more effective than \khedit{simply telling learners whether a response is correct or incorrect.}\khdelete{showing students traditional forms of explicit feedback (e.g., directly telling the student whether their response is correct or incorrect)}
For example, when teaching students to perform algebraic transformations, showing students a graphical representation of \khedit{equations they enter can}\khdelete{their equation that changes to reflect each step the student makes to complete the algebraic transformation, may} help students \khedit{immediately \textit{recognize for themselves} whether an equation is likely to be correct or incorrect}\khdelete{self-evaluate that their steps are correct}\cite{wiese2017designing}. \khdelete{This is in contrast to explicit feedback approaches, that directly verify the correctness of the student's work after completing the problem, rather than having students self-assess their work while they are completing it.}} Future work on training for AI-assisted decision-making should further explore the design space of rich, grounded feedback mechanisms that can help humans (learn to) calibrate their reliance. In particular, whereas our study provided explicit feedback in the form of categorical outcomes and workers’ final decisions, future work should explore what forms of grounded feedback could support more effective process-oriented learning in a given context. In complex decision-making settings such as child maltreatment screening, observable outcomes and even final decisions) are noisy signals for decision making quality~\cite{kawakami2022care,cheng2022child}. Therefore, rather than providing human decision-makers with outcome-based feedback that attempt to directly \khedit{``tell'' a learner whether a given decision was accurate}\khdelete{signal decision accuracy}, providing rich, grounded feedback may make it easier \khedit{to engage learners' sensemaking and help them}\khdelete{for people to} \textit{assess for themselves} whether and why a decision may be right or wrong. Training that focuses on better scaffolding the decision-making process---for example, by pairing a novice decision-maker with an expert decision-maker to collaboratively reason about each decision---may improve critical thinking and sensemaking processes that also allow humans to better calibrate reliance on ADS outputs. 

\akdelete{Interestingly, we also found that participants who receive explicit feedback on their decisions during training \mldelete{recalled}\mledit{referenced} a greater number of information sources (including allegation information) than participants who received repeated practice alone. There are multiple possible interpretations of this finding. For one, it is possible that receiving explicit feedback–which highlighted disagreements between experienced workers' decisions versus the model predictions and proxy outcomes–may have encouraged participants to reflect more on possible specific reasons for such disagreement. As a result of this increased reflection, participants may have explicitly referenced these reasons (e.g., information contained in allegations) more often than those who do not receive feedback. Another possible reason is that participants who received explicit feedback slowed their decision-making process, leading to typing longer explanations for their decisions, which in turn increased the likelihood of recalling more information sources. Future work is needed to understand which hypotheses, and other potential reasons, best explain these results. }

\subsection{On the absence of domain expertise effects} \label{discussion_domainexpertise}
We found that regardless of participants' level of self-reported domain expertise in social work, they learned to disagree with AI predictions in a manner that resembled the disagreement patterns of experienced workers. Participants with greater self-reported domain expertise did not disagree more with the AI prediction or make more accurate decisions with regards to the past experienced worker, 
compared with other participants. 

We do not view the absence of domain expertise effects in our study as evidence that domain expertise has zero impact on how people learn to make AI-assisted decisions. In our study, ``domain expertise'' was broadly defined to include individuals with domain knowledge in \textit{any} area of social work, not limited to knowledge of child welfare or experience with child maltreatment screening. Participants' roles included current or former social workers and social work graduate students. We had originally hypothesized that broad knowledge in social work would influence how and what participants learn through a training for AI-assisted child maltreatment screening. However, given the specialized nature of this task, it is very plausible that a tighter recruitment criterion---for example, requiring ``domain expert'' participants to have prior domain knowledge in child welfare or even prior experience with child maltreatment screening---would reveal more substantial differences between participants with more or less domain expertise. In addition, it is possible that the specific metrics used in our analyses were unable to capture differences in learning and decision-making between the ``domain experts'' in our study and other participants. 
Finally, a third possibility is that general human abilities shared by non-domain-experts (i.e., having access to and being able to interpret qualitative information, and having knowledge of broader decision goals) are sufficient on their own to support learners in approximating experienced workers patterns of disagreement with AI predictions. 

We emphasize that our results do not indicate that participants learned to make decisions \textit{as well as} experienced workers during the course of our training activity. Rather, our findings indicate that participants learned to predict how the AI model would behave on different cases and learned to disagree with the AI in a manner resembling experienced workers. 
Future work is needed to tease apart the three potential explanations above.

\subsection{Generalizability of design decisions and findings}\label{discussion_generalize}
As a concept, \textit{critical use} is applicable to any domain where \khedit{humans and AI models have access to complementary knowledge.}\khdelete{there are target-construct mismatch and information asymmetries.} 
Critical use emphasizes that 
humans can be better supported in calibrating reliance on AI-based decision-making tools, by learning to leverage complementary knowledge they have as human decision-makers (e.g., knowledge of additional decision-relevant features, or knowledge of the true objectives of a decision-making task). 
We expect that several of the findings and implications discussed above---for example, the importance of providing learners with concrete, domain-grounded forms of feedback against which they can cross-check AI outputs---will generalize to other domains, beyond the context studied in the current research. 
However, we note that the indicators of critical use that we adopted in our study are highly specific to the child maltreatment screening context. As discussed above, our choices of indicators were informed by prior empirical research studying how experienced workers in this domain calibrate their reliance upon AI predictions; this specific set of indicators and the interpretations adopted in this study may not generalize well to other domains. Accordingly, our goal in this research is not to make any universal recommendations on which specific, measurable decision-making outcomes and learning indicators are ``good'' or ``bad'' across domains. Rather, we emphasize that 
the approach of drawing upon prior knowledge to design domain-specific training interfaces and indicators of critical use can be generalized to inform the design of improved training materials and learning measures in other complex, real-world domains. 
Future work is need to explore how our approach can be adapted to other domains, and to investigate what training designs and measures of critical use are most appropriate in different decision-making settings. 


\akdelete{
In this study, we designed an experimental study to explore whether \textit{\textbf{repeated practice}} and \textbf{\textit{explicit feedback}} can support humans in making AI-assisted screening decisions towards \textbf{\textit{critical use}}: helping humans learn to effectively calibrate their reliance on AI predictions, by drawing upon their complementary knowledge and abilities as human experts. 
We find that participants, through \textit{repeated practice alone,} learned to make decisions that \textit{disagreed with the AI prediction}. Early on, participants were agreeing with the AI prediction even for cases in which the AI model was inaccurate in predicting the outcome proxy it was trained upon. However, by the end of the training experiment, they disagreed with the AI prediction \textit{as often for cases that were outcome aligned and misaligned}, leading to an overall decreasing trend in agreement with the AI prediction. Interestingly, rather than aligning with the AI prediction, we find that participants were instead learning to \textit{make decisions like past experienced workers} who encountered the same case information (allegations, AI prediction, other case details) that the participants had seen. Therefore, we find that participants learned to make decisions like the past experienced worker, without receiving any explicit feedback that indicated the past workers’ decisions. 


The findings suggest that participants learned to \kholdedit{emulate}\kholddelete{simulate} the behavior of past experienced workers rapidly through repeated practice alone. Past studies analyzing retrospective administrative data recording worker-AFST decsisions~\cite{De-Arteaga2020,cheng2022child} demonstrated how experienced child welfare workers had learned to critically use the AFST, ensuring they were not over-relying on its predictions. While these prior studies examined months and years of historical decision-making, our findings suggest that participants learned to better make decisions with respect to past experienced workers within a remarkably short amount of time. Overall, the findings at least suggest that the training experiment led to improved critical use, by rapidly \textit{decreasing over-reliance} on the AI model. Does this also mean that participants learned to calibrate appropriate reliance on the ADS? The study results indicate that, within the measures that we can define given the data that is available, participants may actually be disagreeing with the AI prediction more often than past experienced workers and more often than what ADS model designers may desire. The findings suggested that participants, with increased practice, learned to agree with the past experienced worker \textit{less} when they had actually agreed with the AI prediction. Moreover, participants also agreed with the past experienced worker, even for cases in which their decisions turned out to be misaligned with the AI model’s outcome target. However, given that the AI model’s target is an imperfect and, in this case, a misaligned proxy for the participants’ true outcomes of interest, there is reason to hold caution in forming normative conclusions from these findings. While these findings shed light onto how workers learned to decrease over-reliance, future work should explore whether and how training approaches can also help \textit{decrease under-reliance}, towards appropriately calibrating reliance. 

Combining \textit{repeated practice with explicit feedback}, including observed outcomes of the proxy that the AI model is trained upon, did not lead to participants disagreeing with the AI prediction more compared to those who received repeated feedback alone. However, it did improve participants’ ability to predict the outcomes of the AI prediction. Importantly, participants in the \textit{Practice + Explicit Feedback} condition were comparatively more likely to learn to make decisions similar to past experienced workers than participants who received repeated practice alone. This suggests that participants are learning from \textit{a subset of the six imperfect proxies} shown as explicit feedback. In particular, it appears that participants learn from feedback on the supervisor’s decision (i.e., the past experienced worker upon which the alignment measure was defined) but not on the AI prediction’s observed outcome (i.e., placement out of home). 


Given that participants learned to make decisions like past experienced workers even without explicit feedback, does this mean that feedback is not as helpful in these social decision-making settings? An examination of the open text responses indicates that some of the learning can be understood by shedding light on the decision-making\textit{ process} that participants engaged in. In particular, participants justified their AI-assisted decisions by drawing upon rich, qualitative information not accessible to the AI model. Sometimes, they used this qualitative information to explain why they believed the AI model was incorrect. On the other hand, participants relied on using the AI prediction to explain their decisions less in the post-assessment compared to the pre-assessment. Moreover, their decision-making goals, as described in open-text responses at the end of the study, indicate that their targets for child maltreatment screening are often heavily misaligned with the target of the AI model while more aligned with the target of the past experienced worker (e.g., focusing on near-term, rather than long-term outcomes). However, these findings do not point to feedback being unhelpful. While the findings suggest the feedback in the study had an effect on participants’ alignment with the past experienced worker, these open text responses also suggest that participants may benefit from receiving \textit{process-oriented feedback}, not just outcome-based feedback. For example, feedback that includes past workers’ open-text case notes (e.g., studied in \cite{saxena2022unpacking}), which documents their \textit{case-specific reasons} for screening in or out a given referral, may be a critical source of learning for participants. Given the contested nature of using outcome-based ``accuracy’’ proxies for AI-assisted social decision-making, providing qualitative, case-specific reasoning that sheds light on \textit{how} decisions were made may be a promising avenue for exploring the design of future training materials. 

Importantly, this study was conducted with an assumption that the AI model and human decision-maker may indeed have\textit{ complementary }strengths and weaknesses, such that a goal towards learning towards critical use of the AI model is appropriate. However, our findings surface potential challenges with this assumption. In our study, participants may be learning to disagree with the AI prediction by learning that certain feedback signals (e.g., observed outcomes on placement within two years) do not align with their own perceptions of what may be meaningful decision outcomes and goals. Grounding interpretations of the feedback signals and AI prediction with the qualitative allegations may have given participants the opportunity to reflect on whether the feedback outcomes and AI model could complement, or even align with their own decision-making goals. Such behaviors and realizations of misaligned goals with the ADS model are reflected in the experiences of actual  child maltreatment screeners using ADS tools in the real-world~\cite{kawakami2022improving}. Moreover, these challenges raise broader questions about the justifiability and validity of using AI-based predictive tools in \textit{social decision-making settings}, where decisions may rely upon inferences and predictions about the intentions or behaviors of other people~\cite{barocas-hardt-narayanan,kawakami2022care, lee2013social}.




Finally, in this study, there were no significant differences in learning between participants with and without domain knowledge in social work. However, this does not necessarily mean that human decision-makers with relevant domain knowledge do not learn more quickly or effectively than someone more isolated from the setting. Those with domain knowledge included participants working broadly in social work---not specifically in child services---and had differing levels of in-practice experience, given some were graduate students while others were practicing social workers. Although we could not use any models to predict learning differences within the participants with social work domain knowledge (given there were not enough participants to assign to different groups), future work is needed to better understand this result. 


}

\begin{acks}
\akedit{We acknowledge support from Toyota Research Institute (TRI), the UL Research Institutes through the Center for Advancing Safety of Machine Intelligence (CASMI) at Northwestern University, the Carnegie Mellon University Block Center for Technology and Society under Award No. 53680.1.5007718 and 55410.1.5007719, the National Science Foundation (NSF) under Award No. 1939606, 2001851, 2000782, and 1952085, and two awards from the National Science Foundation Graduate Research Fellowship Program (GRFP).}
\end{acks}

\akdelete{\section{Conclusion} \label{Conclusion}
AI-assisted decision-making is typically studied in the context of controlled laboratory studies. While necessary to isolate factors of interest, these studies circumvent the messy realities of real-world decision-making tasks, making assumptions about a golden-standard "ground truth" and uniformly distributed information. In this work, we investigated how people with different levels of domain knowledge learn, through practice and feedback, to calibrate their reliance on AI predictions in a real-world setting in which a single, agreed upon “ground truth” is unavailable and information is shared asymmetrically. Our findings were surprising: participants, even those with no a priori domain knowledge, learned to make decisions in ways that aligned more with past experienced workers even in the absence of any explicit feedback on past workers’ decisions. This suggests that training can help mitigate over-reliance on AI models. However, our work also suggests that workers with more experience may take this too far, ultimately placing too little emphasis on feedback from AI models. In future work, we intend to explore how our training approaches could be adapted to help decision-makers find the optimal balance between their own knowledge-driven intuition and recommendations from AI models.}

\bibliographystyle{ACM-Reference-Format}
\bibliography{references}

\end{document}